\documentclass[12pt]{iopart}

\usepackage{amssymb}
\usepackage{graphicx}
\usepackage{url}
\usepackage{units}
\usepackage{upgreek}
\usepackage{orcidlink}

\newcommand*{\finesse}{\mathcal{F}}

\begin{document}

\title[]{Laser-cut Patterned, Micrometer-thin Diamond Membranes with Coherent Color Centers for Open Microcavities}


\author{Yanik~Herrmann$^{1,\diamondsuit}$\orcidlink{0000-0003-0545-7002}, Julia~M.~Brevoord$^{1,\diamondsuit}$\orcidlink{0000-0002-8801-9616}, Julius~Fischer$^{1,\diamondsuit}$\orcidlink{0000-0003-2219-091X}, Stijn~Scheijen$^1$\orcidlink{0009-0009-4092-4674}, Colin~Sauerzapf$^{1,3}$\orcidlink{0009-0004-7655-9289}, Nina~Codreanu$^1$\orcidlink{0009-0006-6646-8396}, Leonardo~G.~C.~Wienhoven$^1$\orcidlink{0009-0009-7745-3765}, Yuran~M.~Q.~van~der~Graaf$^1$\orcidlink{0009-0004-3730-8815}, Cornelis~F.~J.~Wolfs$^1$\orcidlink{0009-0007-1776-3441}, R\'{e}gis~M\'{e}jard$^1$\orcidlink{0009-0008-0181-6163}, 
Maximilian~Ruf$^{1,4}$\orcidlink{0000-0001-9116-6214},
Nick~de~Jong$^2$\orcidlink{0009-0001-3033-252X}, Ronald~Hanson$^{1,*}$\orcidlink{0000-0001-8938-2137}}
\address{$^1$ QuTech and Kavli Institute of Nanoscience, Delft University of Technology, P.O. Box 5046, 2628 CJ Delft, The Netherlands}
\address{$^2$ Netherlands Organisation for Applied Scientific Research (TNO), P.O. Box 155, 2600 AD Delft, The Netherlands}
\address{$^3$ Present address: 3rd Institute of Physics, University of Stuttgart, 70569 Stuttgart, Germany}
\address{$^4$ Present address: SandboxAQ, Palo Alto, California, USA}
\address{$^\diamondsuit$ These authors contributed equally to this work.}
\address{$^*$ Author to whom any correspondence should be addressed.}
\ead{r.hanson@tudelft.nl}

\begin{abstract}
Micrometer-scale thin diamond devices are key components for various quantum sensing and networking experiments, including the integration of color centers into optical microcavities. In this work, we introduce a laser-cutting method for patterning microdevices from millimeter-sized diamond membranes. The method can be used to fabricate devices with micrometer thicknesses and edge lengths of typically \unit[10]{\textmu m} to \unit[100]{\textmu m}. We compare this method with an established nanofabrication process based on electron-beam lithography, a two-step transfer pattern utilizing a silicon nitride hard mask material, and reactive ion etching. Microdevices fabricated using both methods are bonded to a cavity Bragg mirror and characterized using scanning cavity microscopy. We record two-dimensional cavity finesse maps over the devices, revealing insights about the variation in diamond thickness, surface quality, and strain. The scans demonstrate that devices fabricated by laser-cutting exhibit similar properties to devices obtained by the conventional method. Finally, we show that the devices host optically coherent Tin- and Nitrogen-Vacancy centers suitable for applications in quantum networking.
\end{abstract}



\maketitle

\section{Introduction}

Single-crystal (sub) micrometer-thin diamond samples hosting coherent color centers are relevant for several quantum technology applications spanning from quantum networking to sensing. In quantum networking, such platforms can be used to engineer the photonic environment of color centers for enhanced photon collection~\cite{janitz_cavity_2020,mi_integrated_2020,ruf_quantum_2021,shandilya_diamond_2022}. This includes photonic crystal cavities, fabricated directly into the diamond~\cite{riedrich-moller_one-_2012,lee_fabrication_2013,faraon_quantum_2013,ding_high-q_2024} and thinned-down membranes, coated from both sides with a dielectric layer, which can function as an optical antenna~\cite{fuchs_cavity-based_2021}. Furthermore, such platforms can be incorporated into open microcavities~\cite{janitz_fabry-perot_2015,bogdanovic_design_2017,riedel_deterministic_2017} and they facilitate the manufacturing of heterogeneous photonic structures such as solid immersion lenses~\cite{riedel_low-loss_2014}, nanophotonic resonators~\cite{butcher_high-q_2020,guo_direct-bonded_2024} or plasmonic nanogap cavities~\cite{boyce_plasmonic_2024}.\\
In quantum sensing, chemically inert and bio-compatible, thin diamond samples can bring color centers near other materials, while still providing optical access. This can be used in sensing of living cells~\cite{guo_direct-bonded_2024}, wide-field microscopy of electrical currents~\cite{asif_diamond_2024}, or magnetic fields~\cite{Schlussel_wide-field_2018,carmiggelt_broadband_2023,ghiasi_nitrogen-vacancy_2023,borst_observation_2023,zhou_quantum_2023}.\\
In this work, our main focus is on the fabrication of diamond microdevices for open microcavities to Purcell-enhance color centers hosted inside the device. While pioneering work made use of color centers in nanodiamonds~\cite{albrecht_coupling_2013,kaupp_scaling_2013,johnson_tunable_2015}, a micrometer-thin diamond sample is beneficial to maintain good optical coherence, especially in the case of Nitrogen-Vacancy centers~\cite{ruf_optically_2019,yurgens_spectrally_2022}. Such diamond samples have been used to couple Nitrogen-Vacancy and group-IV vacancy centers to open microcavities~\cite{yurgens_cavity-assisted_2024,zifkin_lifetime_2024,berghaus_cavity-enhanced_2025,herrmann_coherent_2024}. The coupling of color centers to the cavity is quantified by the Purcell factor, which is proportional to $Q/V$, where $Q$ denotes the cavity quality factor and $V$ is the cavity mode volume. To demonstrate significant Purcell enhancement of color centers, the following microdevice properties must be taken into consideration: (1) A thickness of a few micrometers (or less) is desired to minimize the cavity mode volume. (2) A smooth surface (roughness $R_q\lessapprox\unit[1]{nm} $) is needed to maintain a high cavity quality factor (finesse). (3) Large lateral dimensions (tens of micrometers) with a small wedge are beneficial to provide an area to probe several cavity spots (cavity beam waist is on the order of a micrometer). (4) Bonding to a cavity Bragg mirror at a micrometer distance to a stripline for delivering microwaves~\cite{bogdanovic_robust_2017} or static electric fields~\cite{tamarat_stark_2006} is necessary. (5) The sample should contain color centers with good optical properties for quantum science and technology applications. While these criteria are relevant for the microcavity platform, they are also relevant to most of the applications mentioned.\\
Several fabrication methods have been developed to realize (sub) micrometer-thin diamond samples. These include the fabrication methods listed in the following: smart cut makes use of the implantation of light ions (like Carbon~\cite{parikh_single-crystal_1992} or Helium~\cite{piracha_scalable_2016-1}) to produce a sacrificial layer of amorphous carbon, which can be removed electrochemically~\cite{fairchild_fabrication_2008,lee_fabrication_2013}. Additional overgrowth is necessary for a high crystalline quality thin film diamond released from the sacrificial layer~\cite{guo_tunable_2021}. In a diamond-on-insulator approach, a thin diamond membrane is bonded to a host (e.g. silicon) substrate and subsequently thinned down to the required thickness~\cite{faraon_resonant_2011,riedrich-moller_one-_2012,hausmann_integrated_2012,ovartchaiyapong_high_2012,faraon_quantum_2013}. Furthermore, $\sim\unit[50]{\upmu m}$ diamond microdevices can be patterned in a one-step (e.g. with a hydrogen silsesquioxane~(HSQ) mask~\cite{appel_fabrication_2016}) or two-step pattern transfer process (e.g. with a silicon nitride ($\rm Si_xN_y$) hard mask~\cite{ruf_cavity-enhanced_2021}) and thinned down by dry etching and a heterogeneously integrated mask (e.g. a fused quartz mask~\cite{tao_single-crystal_2014}). Alternatively, such diamond membranes can be first bonded to the cavity mirror and then thinned down with a fused quartz~\cite{ruf_optically_2019} or diamond~\cite{heupel_fabrication_2023} mask. Other methods start from bulk diamond and use an undercutting process (via angled plasma etching~\cite{burek_free-standing_2012} or quasi-isotropic plasma etching~\cite{khanaliloo_high-qv_2015}). Moreover, the production of free-standing diamond nanoslabs with a Chromium protection mask and dry etching has been demonstrated~\cite{hodges_long-lived_2012}. These fabrication methods are typically complex, labor-intensive, and necessitate additional equipment for the pattern transfer, lithography tools, and deposition tools to define microstructures in lateral dimensions.\\ 
In this work, we present a fabrication method for patterning microdevices of different sizes and shapes using laser-cutting to engrave $\sim\unit[50]{\upmu m}$ thick diamond membranes. This approach utilizes only two fabrication tools (a femtosecond pulsed laser writer and a reactive ion etcher), significantly simplifying the process, reducing its duration, and increasing accessibility. The method allows to achieve feature sizes of approximately \unit[3]{$\upmu$m}, limited by the laser spot size. We compare this new fabrication method with an already established method of patterning the diamond membrane with electron-beam lithography (EBL) based on two-step patterning: the desired design is EBL patterned into a resist, followed by a reactive ion etch (RIE) to transfer the pattern into a hard mask material. Next, the hard mask pattern is transferred into the diamond membrane via RIE~\cite{ruf_cavity-enhanced_2021}. Diamond microdevices obtained from both fabrication methods are then subsequently bonded to a cavity Bragg sample mirror, equipped with striplines for microwave delivery, see Fig.~\ref{fig:scheme}~(a) and (b). To demonstrate that the laser-cut microdevices are suitable for microcavity applications, we characterize the microdevices by scanning cavity microscopy~(SCM)~\cite{toninelli_scanning_2010,mader_scanning_2015,benedikter_transverse-mode_2019}, see Fig.~\ref{fig:scheme}~(d). This demonstrates that both methods lead to microdevices with comparable cavity quality factors (cavity finesses). Furthermore, we study the optical properties of negatively charged Tin-Vacancy (SnV) and negatively charged Nitrogen-Vacancy (NV) centers embedded in diamond microdevices at low temperatures. Both methods result in devices suitable for microcavity applications in quantum networking.

\section{Fabrication of diamond microdevices}
The two fabrication methods are compared and described in detail in the following. The process flow of both methods can be divided into five steps: sample preparation, patterning, color center creation, device release etch, and bonding. Specifically, the two methods majorly differ in the patterning fabrication step, where the method presented here foresees a design pattern transfer via laser-cutting. In contrast, the conventional fabrication method foresees a two-step transfer pattern based on EBL. The laser-cut method with SnV center creation is schematically summarized in Fig.~\ref{fig:scheme}~(c). The conventional EBL-based fabrication method is detailed in~\cite{ruf_cavity-enhanced_2021}. Detailed fabrication steps and parameters can be found in the Appendix Tab.~\ref{tab:fab}.

\subsection{Sample preparation}
\begin{figure}[ht]
    \centering
    \includegraphics[width=\linewidth]{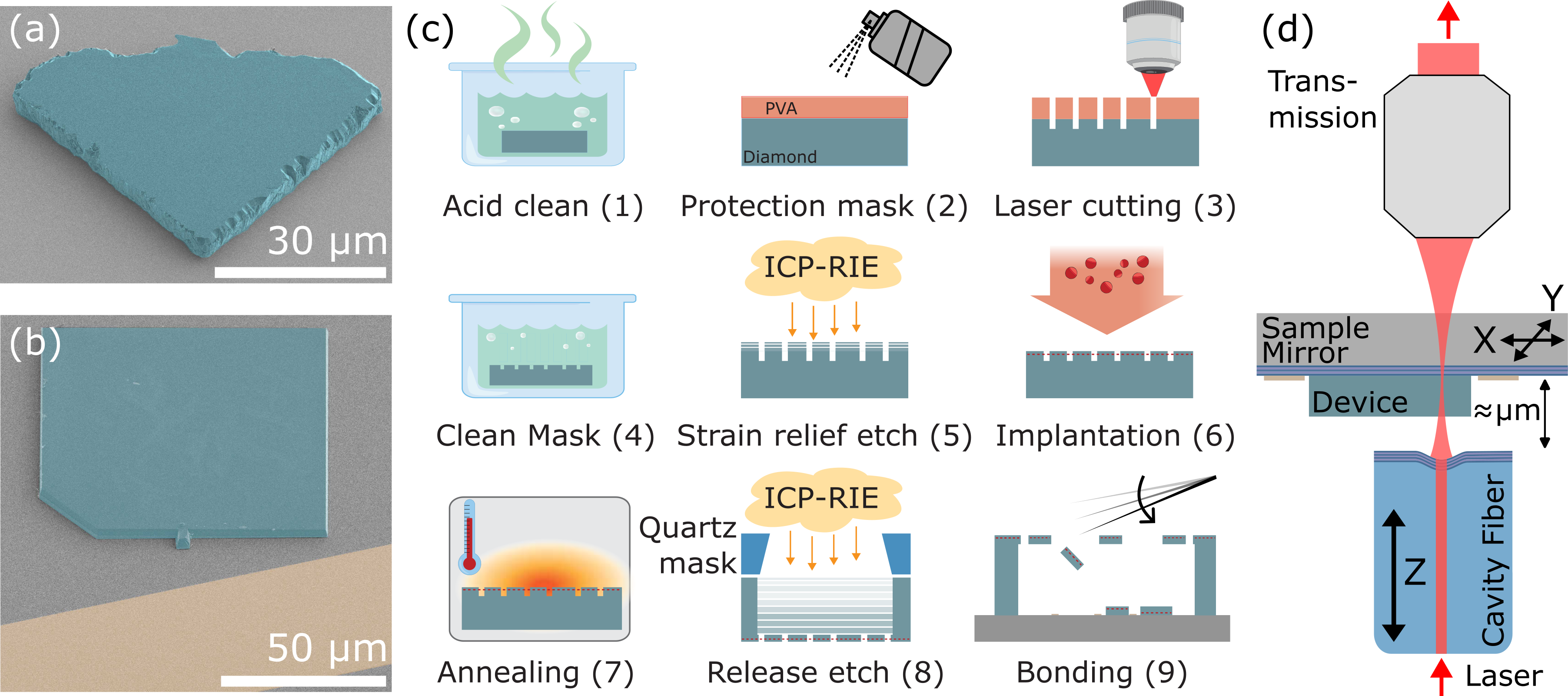}
    \caption{False-colored (cyan) scanning electron microscope (SEM) images of diamond microdevices obtained by laser-cutting (device originated from a parent diamond membrane named: \textit{Vincent Vega}) in (a) and electron-beam lithography, and dry etching with a silicon nitride hard mask (\textit{Pai Mei}) in (b). The devices are bonded to the cavity mirror. A gold stripline (yellow false-colored) can be used for microwave delivery. (c) Process flow to fabricate laser-cut diamond microdevices with Tin-Vacancy centers bonded to the cavity mirror. PVA:  polyvinyl alcohol, ICP-RIE: inductively coupled plasma reactive ion etching. (d)~Schematic of the scanning cavity microscope. The cavity is optically probed from the fiber side, and the transmission is detected. The sample mirror with the bonded diamond microdevices is scanned laterally in front of the fiber.} 
    \label{fig:scheme}
\end{figure}

For both methods, we follow the sample preparation step outlined in Ref.~\cite{ruf_optically_2019}. We start with commercially available single-crystal, electronic-grade bulk diamonds, measuring $\unit[(2 \times 2)]{mm^2}$ and $\unit[0.5]{mm}$ in thickness with a face-orientation of $\left<100\right>$, grown by chemical vapor deposition. The diamonds are laser-sliced into three $\sim\unit[50]{\upmu m}$ thick membranes and polished on both sides (Almax~Easylab). The roughness of the polished surface is typically $R_q<\unit[1]{nm}$. Before patterning, the membranes are cleaned by fuming nitric acid (65~\%) at room temperature (step (1) in Fig.~\ref{fig:scheme}~(c)).

\subsection{Patterning}
\label{sec:patterning}
Patterns of arbitrary microdevice designs are transferred into the $\sim\unit[50]{\upmu m}$ thick diamond membrane via the two methods by selectively removing diamond material.\\
In the case of the laser-cutting method, the diamond membrane is coated with a polyvinyl alcohol (PVA) mask using standard PVA-based hairspray (step (2)). Next, approximately \unit[10]{\textmu m} to \unit[15]{\textmu m} deep trenches are created with a femtosecond-pulsed laser (Lasertec), transferring an arbitrary design into the diamond membrane (step (3))~\cite{eaton_quantum_2019}. After patterning, the PVA mask is removed via an ultrasonic bath in de-ionized (DI) water ($\rm H_2O$) and acetone for \unit[10]{min} each at room temperature (step (4)), followed by inorganic wet cleaning in a Piranha mixture (ratio 3:1 of $\rm H_2SO_4$~(95~\%)~:~$\rm H_2O_2$~(31~\%)) at 80~$^{\circ}$C for \unit[20]{min}. In the next step, the surface is treated with a strain relief etch for cleaning and to remove the residual polishing-induced strain from the membrane surface (step (5))~\cite{appel_fabrication_2016}. The membrane is mounted with PMMA495~A4 on a fused quartz carrier wafer and etched with inductively coupled plasma reactive ion etching (ICP/RIE) for \unit[35]{min} in Ar/$\rm Cl_2$ plasma chemistry, followed by \unit[7]{min} etch in $\rm O_2$ plasma chemistry (Oxford~Instruments, Plasmalab~System~100). This removes $\sim\unit[1.4]{\upmu m}$ and $\sim\unit[2.1]{\upmu m}$, respectively, of diamond material from the surface. Alternatively, the strain relief etch can also be performed with a cyclic Ar/$\rm Cl_2$ and $\rm O_2$ recipe to improve surface roughness, as demonstrated in Ref.~\cite{heupel_fabrication_2023}.\\
The EBL method starts with processing the diamond membrane with a strain relief etch as described above (with different etching times). After the color center creation (described below), a $\sim\unit[320]{nm}$ thin $\rm Si_xN_y$ layer hard mask material is deposited on the top diamond surface by plasma-enhanced chemical vapor deposition (PECVD, Oxford~Instruments Plasmalab~80~Plus). Next, CSAR-13 (AR-P~6200.13) positive tone resist is spin-coated, followed by the spin coating of Electra~92 (AR-PC~5090) conductive polymer to prevent charging effects during EBL exposure (Raith EPBG-5200). Before the development step, the Electra~92 is removed in deionized water and blow-dried with nitrogen. The resist is then developed by immersing the sample for \unit[1]{min} in pentyl acetate, \unit[5]{s} in ortho-xylene, \unit[1]{min} in isopropyl alcohol (IPA), and blow-drying with nitrogen. The pattern is transferred into the $\rm Si_xN_y$ hard mask layer via ICP/RIE (Adixen~AMS~100~I-speeder) in a $\rm CHF_3/O_2$ based plasma chemistry~\cite{norte_mechanical_2016}. The resist is removed in a two-fold step: first, a coarse resist removal is executed by immersing the sample in a PRS~3000 positive resist stripper solution, followed by a second fine resist removal in a Piranha mixture clean, which is executed two times. Next, the pattern is transferred from the $\rm Si_xN_y$ hard mask into the diamond membrane by ICP/RIE with an $\rm O_2$ plasma (Oxford~Instruments, Plasmalab~System~100), resulting in a trench depth of around \unit[6]{\textmu m} to \unit[10]{\textmu m}. The $\rm Si_xN_y$ hard mask is then removed by wet inorganic etch in hydrofluoric (HF) acid (40~\%) for \unit[15]{min} at room temperature.

\subsection{Color center creation}
\label{sec:cccreation}

SnV centers are created in the diamond membrane by ion implantation of Tin ($\rm ^{120}Sn$) at an implantation energy of \unit[350]{keV} and a dose of $\unit[3 \times 10^{10}]{\rm ions/cm^2}$ under an implantation angle of 7$^{\circ}$ (step (6)). The desired implantation energy is determined using Stopping Range of Ions in Matter (SRIM) simulations~\cite{ziegler_srim_2010}, resulting in an implantation depth of $\sim\unit[90]{nm}$ (straggle of $\sim\unit[17]{nm}$), and the implantation angle is adopted to prevent ion channeling effects. For laser-cut samples, ion implantation is performed after patterning and strain relief etch. For EBL samples, the ion implantation is performed after the strain relief etch, but before the patterning step.\\
For the generation of NV centers with high optical coherence, we employ electron irradiation with minimal crystal damage to generate NV centers from the native nitrogen concentration~\cite{ruf_optically_2019,tamarat_stark_2006,van_dam_optical_2019-1}. We irradiate and anneal the bulk diamonds before laser-slicing into membranes, but this step can also be performed later in the process. Irradiation is realized at the Reactor Institute in Delft with an electron-beam acceleration energy of \unit[2]{MeV} and a dose of $\unit[4\times10^{13}]{e^{-}/cm^{2}}$. The energy leads to penetration of electrons through the full diamond (for both substrates used in this study, the \unit[0.5]{mm} thick bulk and $\sim\unit[50]{\upmu m}$ membranes). This results in lattice vacancies formed over the full thickness extent of the diamond substrates.\\
After the implantation and the irradiation steps, the diamond substrates are processed with a tri-acid clean (mixture of 1:1:1 ratio of $\rm H_2SO_4$(97~\%):$\rm HNO_3$(65~\%):$\rm HClO_4$(60~\%)) at 120~$^{\circ}$C for one hour. To enable vacancy migration and to activate color centers, as well as to remove defects (such as divacancies from the lattice), the samples are high-temperature annealed (at a pressure of $\sim\unit[1 \times 10^{-6}]{mBar}$) with a temperature of up to 1100~$^{\circ}$C (step (7)), followed by a tri-acid clean to remove any created graphite during the annealing step~\cite{ruf_optically_2019}. Both ways to create color centers (implantation and electron irradiation) are compatible with the two demonstrated fabrication methods.\\

\subsection{Device release etch}
To release the structures, the membrane with the patterned side facing down is placed on a fused quartz wafer. A \unit[0.5]{mm} thick fused quartz mask with a \unit[1.4]{mm} rectangular central opening and angled sidewalls (30$^{\circ}$) is positioned and aligned on top of the back surface of the diamond membrane~\cite{ruf_optically_2019}. The mask allows the etching plasma to access the back surface of the diamond membrane over the entire opening area, whereas areas of the membrane covered yield a supporting diamond frame. The center part of the membrane is etched with an ICP/RIE process (Oxford~Instruments, Plasmalab~System~100) composed of Ar/$\rm Cl_2$ for \unit[45]{min} for cleaning and smoothening the surface, followed by consecutive multiple rounds of $\rm O_2$ plasma deep etching (step (8), depending on the diamond membrane thickness). Between etching rounds, the membrane is inspected under an optical light microscope to assess the progressing thickness and verify the release of the microdevices, connected to the parent membrane by a small tether by design. The device release etch step is concluded when the target thickness is reached.

\subsection{Bonding}
Prior to the diamond microdevice bonding to the cavity sample mirrors~\cite{riedel_deterministic_2017}, microwave striplines can be fabricated on top of the mirrors. These are produced by an EBL and lift-off fabrication process, resulting in \unit[5]{nm} titanium and \unit[75]{nm} thick, \unit[50]{\textmu m} wide gold striplines. A wet inorganic Piranha clean concludes the microwave stripline fabrication, improving the bonding quality of the diamond microdevices. The stripline fabrication process does not reduce the cavity mirror reflectivity, as demonstrated in Ref.~\cite{bogdanovic_robust_2017}.\\
The bonding is performed by a four-degrees-of-freedom, piezo micromanipulator (Imina~Technologies~SA, miBots), used to controllably break the tether and release the diamond microdevices from the parent membrane (shown in Appendix Fig.~\ref{fig:bonding}~(a), see also Supplementary Video~1). As a result, the microdevices land on the cavity mirror, aligned and positioned under the parent diamond membrane such that proximity to the microwave striplines is ensured (step (9)). The micromanipulator is equipped with a \unit[0.6]{\textmu m} fine tungsten needle (Ted~Pella, Ultrafine~Tungsten Manipulator Probe). To increase the probability of obtaining well-bonded devices, the surface of the cavity mirror can be activated for \unit[30]{min} inside an Ozone chamber (BioForce~Nanosciences UV/Ozone~ProCleaner), which removes organic contamination on the molecular level. In addition, the overall parent diamond membrane with the released microdevices can be cleaned in HF (40~\%) at room temperature. A well-bonded device exhibits a color matching with the mirror surface and does not show any interference patterns. Furthermore, these devices do not move when a lateral force with the micromanipulator tip is applied. Bonded microdevices from both fabrication methods are shown in Fig.~\ref{fig:scheme}~(a) and (b), measured with an SEM, and in Fig.~\ref{fig:classical_microscope}~(a) and (c) with a light microscope.\\
In contrast, devices that are not fully bonded are typically identified by wave-like interference patterns (Newton's rings) or an opaque appearance (see Appendix Fig.~\ref{fig:bonding}~(b)~and~(c)). Unbonded devices can be repositioned on the mirror surface for alignment, such as placement near the stripline. For some devices, gentle force or tapping with the micromanipulator has resulted in bonding. Both fabrication methods, laser-cutting (Fig.~\ref{fig:classical_microscope}~(a)) and EBL (Fig.~\ref{fig:classical_microscope}~(c)), produce well-bonded devices.\\

\subsection{Fabrication results}

\begin{figure}[ht]
    \centering
    \includegraphics[width=0.8\linewidth]{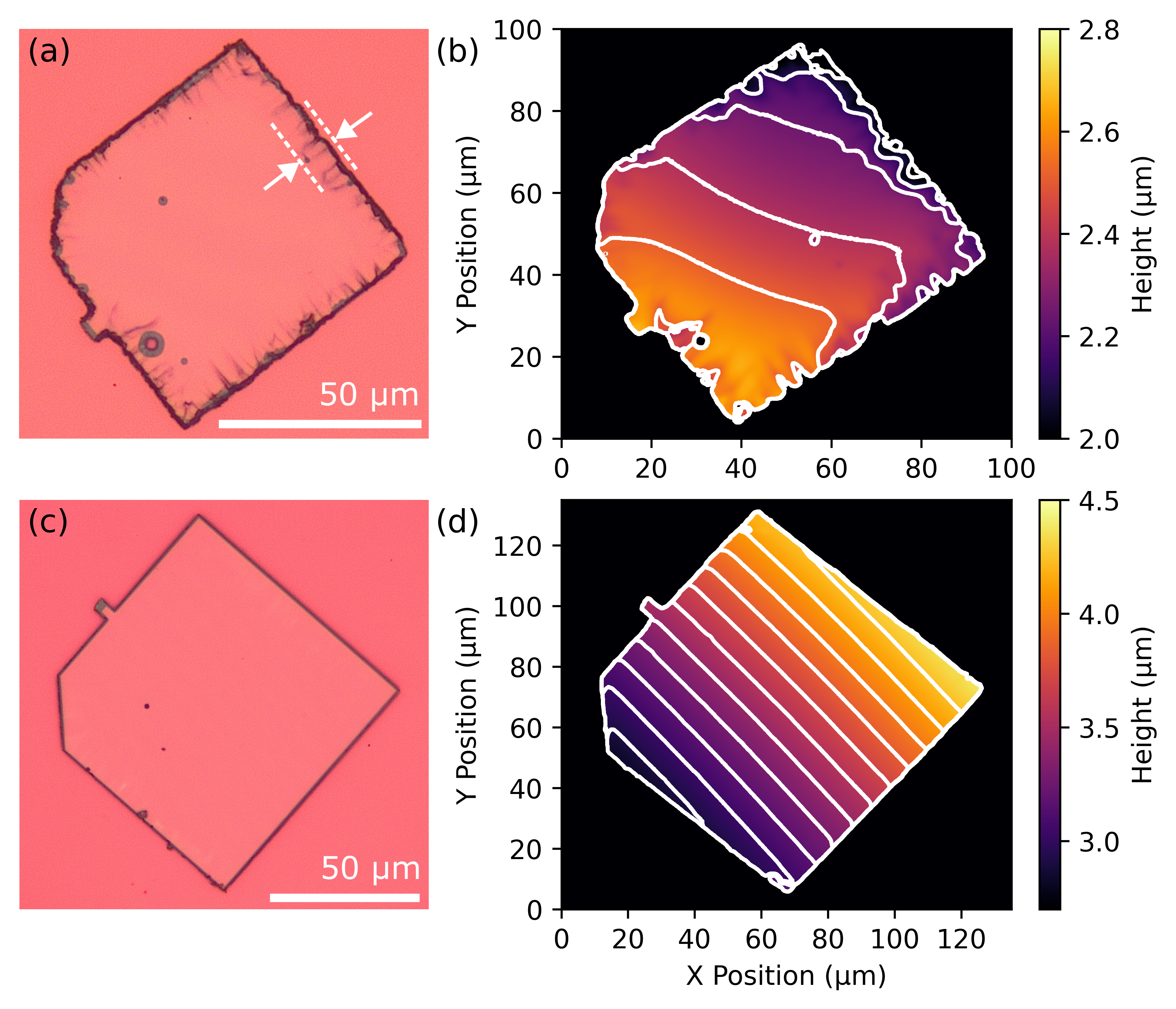}
    \caption{Comparison of diamond microdevices fabricated by laser-cutting and EBL. (a) Optical light microscope picture of the laser-cut microdevice (\textit{Vincent Vega}). The rougher edge (indicated by the \unit[10]{\textmu m} spaced, dashed lines) of the laser-cut microdevice is visible. (b) The corresponding height map is measured with a white light interferometer. The white lines are overlayed and indicate a microdevice thickness matching air-like modes with $q=16$ to $q=19$, calculated by equation~(\ref{equ:thickness_mode}). (c) Optical light microscope picture of the EBL microdevice (\textit{Pai Mei}). (d) Corresponding height map with diamond thicknesses leading to an air-like mode of $q=22$ to $q=32$, calculated by equation~(\ref{equ:thickness_mode}). Both devices are well-bonded to the underlying (pink) cavity mirror. The height data of the white light interferometer is stated with respect to the mirror surface.} 
    \label{fig:classical_microscope}
\end{figure}

Representative examples of bonded microdevices fabricated with the two methods presented in the previous sections are shown in Fig.~\ref{fig:classical_microscope}. A laser-cut \unit[($70 \times 70$)]{\textmu m$^2$}, ${\sim\unit[2.5]{\upmu m}}$ thin microdevice shown in Fig.~\ref{fig:classical_microscope}~(a) hosts SnV centers. The corresponding height map in Fig.~\ref{fig:classical_microscope}~(b) is measured by a white light interferometer (Bruker~ContourX-500) and yields a wedge with a slope of ${\sim\unit[0.7]{\upmu m/100 \upmu m}}$. The second microdevice, fabricated by EBL is \unit[($90 \times 90$)]{\textmu m$^2$} in size and around \unit[3.5]{\textmu m} thin, shown in Fig.~\ref{fig:classical_microscope}~(c) and contains NV centers. The height maps (Fig.\ref{fig:classical_microscope}~(d)) shows a wedge of ${\sim\unit[1.4]{\upmu m/100 \upmu m}}$. The higher gradient wedge indicates that this device originated from the outer region of the parent diamond membrane, where the etched profile is less homogeneous, because of the proximity effect due to the fused quartz mask. The determined wedge of both microdevices enables the investigation of the two different mode types, which are formed in the diamond-microcavity system (air-like and diamond-like, see next section).\\
The laser-cutting method yields devices with a significantly higher surface roughness for an extent of $\sim\unit[10]{\upmu m}$ from the edges when compared to the EBL method obtained devices (compare Fig.~\ref{fig:classical_microscope}~(a)~to~(c)). Beyond the high surface roughness extent of $\sim\unit[10]{\upmu m}$, the inner area of the laser-cut microdevice shows a comparable surface quality to the devices fabricated by EBL.\\
Both methods lead to the successful fabrication of microdevices with high-quality surfaces in the center, suitable for microcavity experiments (next section).\\

\newpage
\section{Scanning cavity microscopy}

In this section, we use SCM (see Supplementary Video~2) to study the quality of the bonded diamond microdevices. The experiments involve finesse measurements of the microcavity that give insights into cavity losses and allow us to conclude the diamond device surface wedge and quality. For this, we first introduce a cavity loss model to describe the finesse determined in length as done in our cavity finesse measurements.

\subsection{Cavity loss model}
We consider a hemispherical plano–concave microcavity, where both Bragg mirrors are terminated with a high refractive index material and the diamond microdevices are bonded to the (flat) sample mirror (see Fig.~\ref{fig:scheme}~(d)). The cavity finesse $\finesse=2\pi/\mathcal{L}_{\rm eff}$ is inversely proportional to the sum of the effective losses $\mathcal{L}_{\rm eff}$. Together with the known mirror transmission loss values, this results in a complete understanding of the system. Most fundamentally, the diamond devices change the sample mirror transmission loss for a given probe laser wavelength depending on their thickness. Hereby so-called air-like and diamond-like mode~\cite{janitz_fabry-perot_2015,van_dam_optimal_2018} thicknesses can be defined: 

\begin{equation}
    \label{equ:thickness_mode}
    t_d = q\frac{\lambda}{2n_d} \quad {\rm(air-like)}, \quad \quad \quad t_d = (2q+1)\frac{\lambda}{4n_d} \quad {\rm(diamond-like)},
\end{equation}

with a corresponding fundamental mode number $q$ and the refractive index of diamond $n_d=2.41$~\cite{zaitsev_optical_2001} (for our laser wavelength of $\lambda=\unit[637]{nm})$. For the air-like (diamond-like) modes, the mirror transmission loss is minimal (maximal). This phenomenon is associated with the electric field intensity ratio of the air and the diamond part of the cavity. A consecutive air-like and diamond-like mode is separated by a diamond thickness of $\lambda/(4n_d)\approx\unit[66]{nm}$ and hence can indicate a device thickness wedge.\\
To include further diamond-related loss mechanisms, we can model the effective losses for the here conducted finesse measurements with:

\begin{equation}
\label{equ:effective_losses}
    \mathcal{L}_{\rm{eff}} = \mathcal{L}_{M,a} + \frac{n_d E_{\rm{max,d}}^2}{E_{\rm{max,a}}^2} \left( \mathcal{L}_{M,d} + \mathcal{L}_{A,d} + \mathcal{L}_{S,\rm{eff},d} \right) + \mathcal{L}_{\rm add}.
\end{equation}

In this equation, $\mathcal{L}_{M,a}$ accounts for the loss through the air-side mirror and $\mathcal{L}_{\rm add}$ considers any additional losses. The fraction ${n_d E_{\rm{max,d}}^2}/{E_{\rm{max,a}}^2}$ describes the electric field intensity ratio between the air and diamond part, and can be calculated by~\cite{van_dam_optimal_2018}:

\begin{equation}
\label{equ:electric_field_intensity_ratio}
    \left( \frac{n_d E_{\rm{max,d}}^2}{E_{\rm{max,a}}^2} \right)^{-1} = \frac{1}{n_d} \sin^2\left( \frac{2 \pi n_d t_d}{\lambda} \right) + n_d \cos^2\left( \frac{2 \pi n_d t_d}{\lambda} \right).
\end{equation}

The ratio depends on the diamond thickness and leads to the modulation of the finesse values between air-like (${n_d E_{\rm{max,d}}^2}/{E_{\rm{max,a}}^2}=1/n_d$) and diamond-like modes (${n_d E_{\rm{max,d}}^2}/{E_{\rm{max,a}}^2}=n_d$). The modulated losses are: the diamond-side sample mirror loss $\mathcal{L}_{M,d}$, the absorption losses $\mathcal{L}_{A,d}$ and the scattering losses at the air-diamond interface $\mathcal{L}_{S,\rm{eff},d}$. The absorption losses can be calculated with the diamond absorption coefficient $\alpha$ to $\mathcal{L}_{A,d}\approx2\alpha t_d$, while the scattering losses can be calculated by~\cite{van_dam_optimal_2018}:

\begin{equation}
\label{equ:scattering_loss}
    \mathcal{L}_{S,\rm{eff},d} \approx \sin^2\left( \frac{2 \pi n_d t_d}{\lambda} \right) 
    \frac{(n_d + 1)(n_d - 1)^2}{n_d}  \left( \frac{4 \pi \sigma_{DA}}{\lambda} \right)^2.
\end{equation}

Here, $\sigma_{DA}$ includes the device surface roughness at the air-diamond interface. For an air-like mode, the scattering losses are zero, while they become maximal in a diamond-like mode. The cavity finesse on the diamond is thus also a measure of the surface roughness.\\

\subsection{Cavity finesse measurements}
\label{sec:finesse_measurements}

\begin{figure}[ht]
    \centering
    \includegraphics[width=0.85\linewidth]{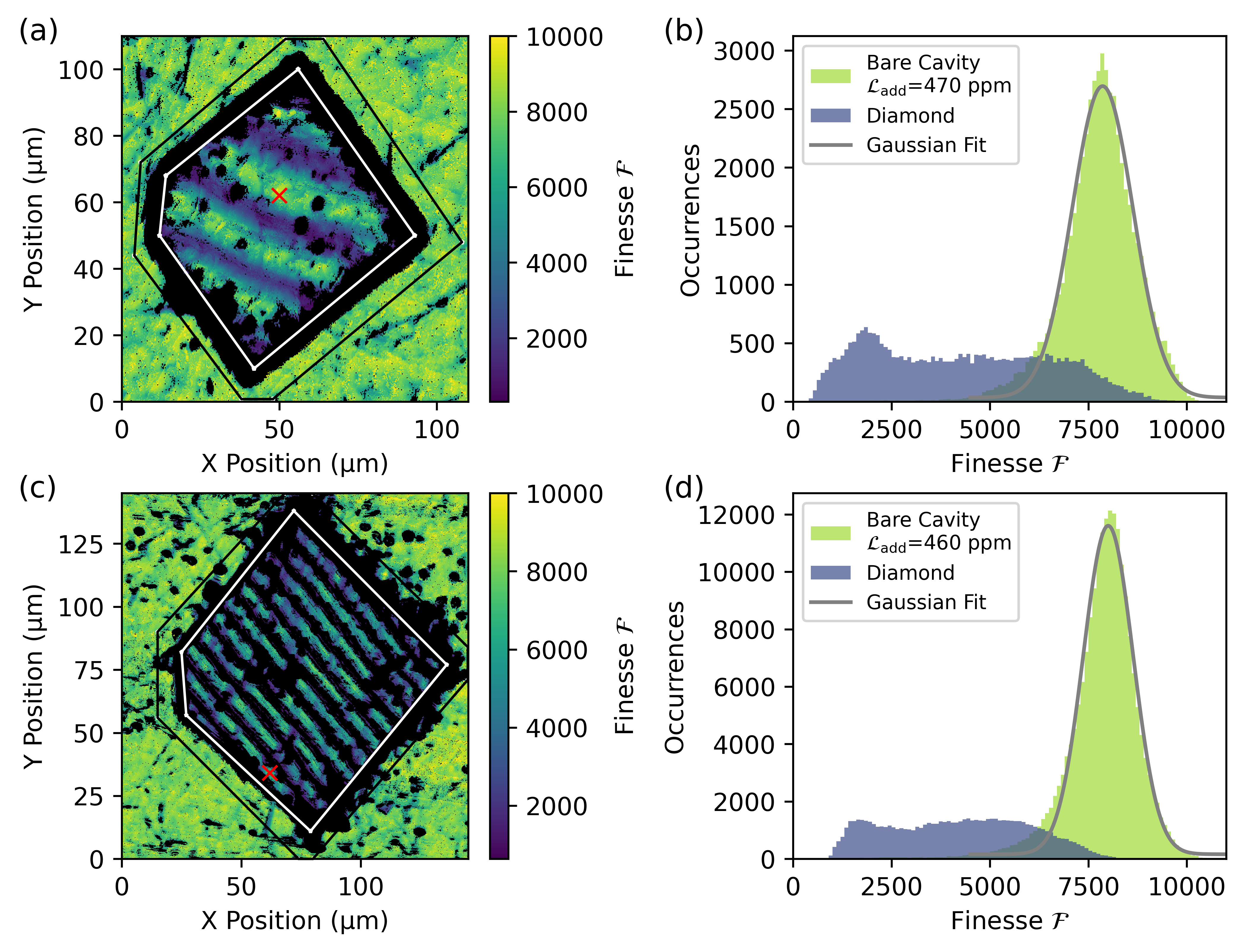}
    \caption{Scanning cavity microscopy of the diamond microdevices. Both devices show a clear modulation of the finesse due to the alternation between air-like and diamond-like modes. (a) Measured two-dimensional finesse scan on a laser-cut diamond microdevice (\textit{Vincent Vega}). (b) Histogram of the finesse values measured in (a). Values on the diamond (bare cavity) are taken from inside the white rectangle (outside the black rectangle). (c) Measured two-dimensional finesse scan on an EBL diamond microdevice (\textit{Pai Mei}). (d) Histogram of finesse values measured in (c). Finesse values on the diamond (bare cavity) are taken from inside the white rectangle (outside the black rectangle). For the bare cavity, a finesse centered around 8000 is determined by the Gaussian fits in (b) and (d). The red crosses in (a) and (c) indicate the positions where the cavity length measurements presented in \ref{app:hybrid} are taken. The histograms are plotted in bins of 100.} 
    \label{fig:scm}
\end{figure}

We characterize the diamond microdevices by two-dimensional scans, in which the cavity finesse is measured on each lateral spot with a room temperature fiber-based Fabry-P\'{e}rot microcavity, schematically depicted in Fig. \ref{fig:scheme}(d) (scanning cavity microscopy~\cite{korber_scanning_2023}). The $\unit[(1 \times 1)]{cm^2}$ sample mirror, on which the diamond devices are bonded, is scanned laterally with a piezo nanopositioning stage in steps of \unit[0.2]{\textmu m}. Technical details about the microcavity setup can be found in \ref{app:setup}. An individual finesse measurement is performed by scanning the cavity length over two fundamental modes while probing the cavity transmission with a resonant laser, shown in Appendix Fig.~\ref{fig:fiber}~(a). We record all measurements with cavity lengths below \unit[15]{\textmu m}, to ensure that the cavity finesse is not limited by clipping losses (see Appendix Fig.~\ref{fig:fiber}~(b) for characterization of the cavity fiber). A scan of the two devices fabricated by laser-cutting and EBL is presented in Fig.~\ref{fig:scm}~(a) and (c).\\
The mirror coatings (Laseroptik) are designed to be maximally reflective for \unit[637]{nm} light with $\mathcal{L}_{M,a}=\unit[50]{ppm}$ ($\mathcal{L}_{M,d}=\unit[670]{ppm}$ for diamond termination) losses for the fiber tip (sample) mirror. As follows from equation~(\ref{equ:effective_losses}), in a bare cavity ($t_d=\unit[0]{\upmu m}$), scattering and absorption losses vanish and the sample mirror transmission losses translate to the air-terminated value of $1/n_d\times\mathcal{L}_{M,d}=\unit[280]{ppm}$. The measured finesse around 8000 (see green distributions in Fig.~\ref{fig:scm}~(b) and (d)) indicates additional losses around $\mathcal{L}_{\rm add}=\unit[470]{ppm}$. We attribute these additional losses to the quality of the concave fiber tip.\\
On both diamond microdevices in Fig.~\ref{fig:scm}~(a) and (c), the finesse shows a clear modulation with the same pattern as in the diamond thickness variation from Fig.~\ref{fig:classical_microscope}~(b) and (d). We associate the variation in finesse values with the alternation between air-like modes (higher finesse) and diamond-like modes (lower finesse). In the air-like modes, only a slight reduction of the cavity finesse compared to the bare cavity is observed. For the diamond-like modes, the finesse is much reduced. The cavity transmission on the EBL device (\textit{Pai Mei}) is reduced around diamond-like mode regions so much that no finesse can be determined. Some thin lines with reduced finesse are visible in the scan in Fig.~\ref{fig:scm}~(c), which are attributed to transverse-mode mixing~\cite{korber_scanning_2023}.\\

\newpage

\begin{figure}[ht]
    \centering
    \includegraphics[width=0.9\linewidth]{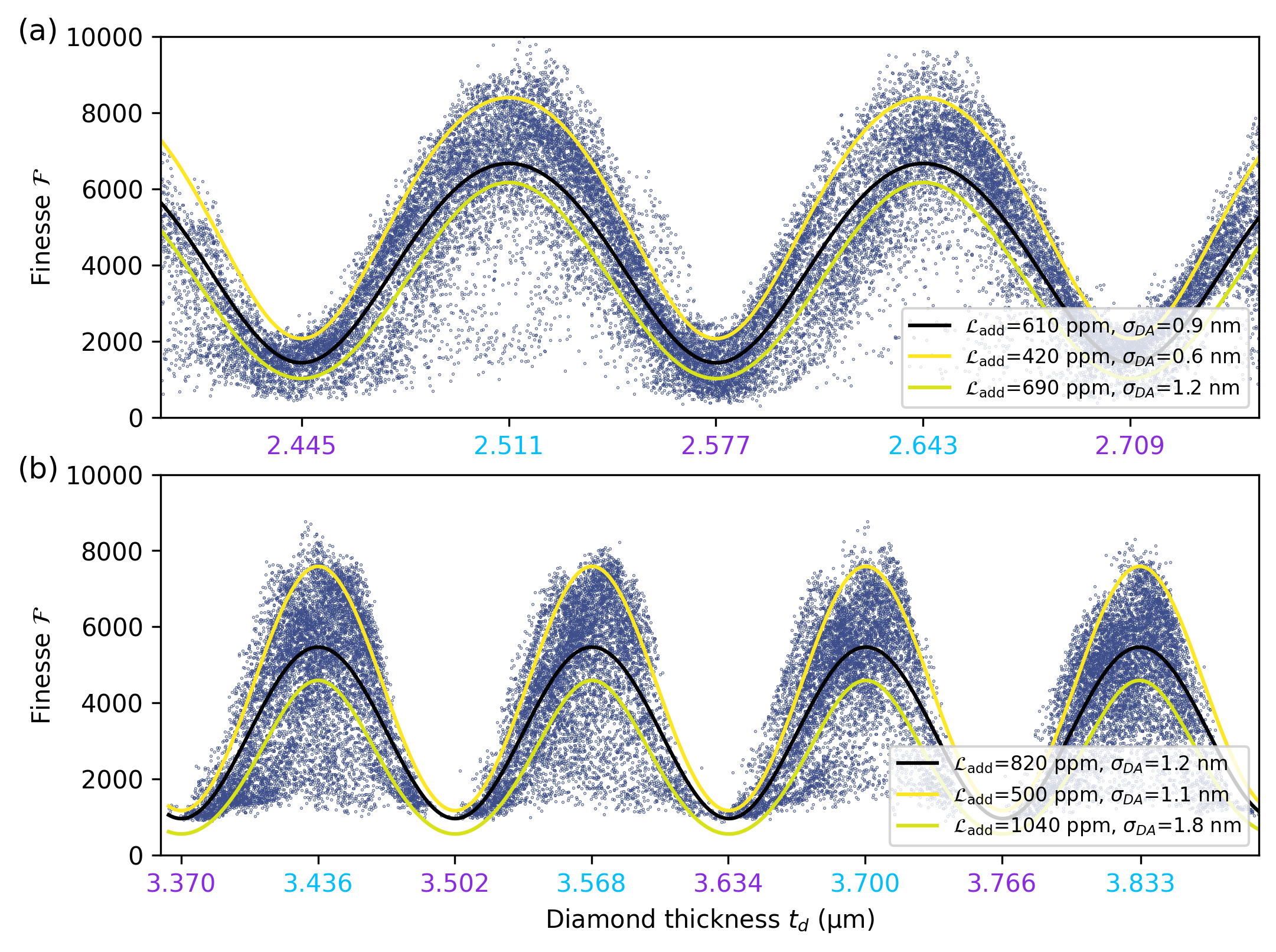}
    \caption{Diamond thickness dependent finesse values observed by SCM. Finesse values are taken from inside the white rectangle in Fig.~\ref{fig:scm}. The black line is a fit of the finesse values with losses from equation~(\ref{equ:effective_losses}). The modulation of the finesse values between high-finesse air-like modes and low-finesse diamond-like modes is visible. The diamond thickness values that correspond to an air-like (diamond-like) mode are indicated by the bright blue (purple) axis ticks, determined by equation~(\ref{equ:thickness_mode}). By determining the mean and standard deviation of the finesse values in \unit[10]{nm} segments (see text), we also give an upper (yellow) and lower finesse curve (bright green) for both devices. More than 60~\% of the finesse values lie between the two curves. The diamond thickness measurement by the cavity dispersion is performed for the displayed first air-like mode for both samples (see Appendix Fig.~\ref{fig:wavy}). (a) Finesse measured on the laser-cut diamond microdevice (\textit{Vincent Vega}). (b) Finesse measured on the EBL diamond microdevice (\textit{Pai Mei}). Note that the black fit is lowered due to the higher spread in finesse values.} 
    \label{fig:finesse_fit}
\end{figure}

\noindent To study the effect of the diamond microdevices more quantitatively, we investigate the finesse depending on the device thickness. This data is obtained by overlaying the white light interferometer data from Fig.~\ref{fig:classical_microscope}~(b) and (d) with the finesse scans from Fig.~\ref{fig:scm}~(a) and (c). The interferometer data has a sufficient lateral resolution of about \unit[0.13]{\textmu m} per step. For every measured finesse value, within the white rectangle in the scan, we take the corresponding diamond thickness from the interferometer data. This is shown in Fig.~\ref{fig:finesse_fit} for both diamond devices.\\
We notice that the absolute thickness values from the white light interferometer do not directly match the expected positions of the air-like and diamond-like modes according to equation~(\ref{equ:thickness_mode}). The measured values might be distorted due to the multilayer Bragg mirror on which the devices are bonded. We compensate for this by measuring the diamond thickness using the cavity mode dispersion (Appendix Fig.~\ref {fig:wavy}) on an air-like mode on the diamond (indicated in Fig.~\ref{fig:scm}) to determine an offset for the diamond thickness values. The absolute thickness from the white light interferometer and the thickness determined by the cavity dispersion differ by \unit[0.16]{\textmu m} to \unit[0.32]{\textmu m}.\\
We fit the measured finesse values depending on the diamond thickness in Fig.~\ref{fig:finesse_fit} with the effective losses from equation~(\ref{equ:effective_losses}). In the fit, we use the known mirror losses from the coating design as stated above, and a small diamond thickness translation of $<\unit[20]{nm}$ is allowed. Furthermore, for the used high-purity, electronic-grade diamonds grown by chemical vapor deposition, no absorption losses are expected~\cite{friel_development_2010}. This leaves the additional losses $\mathcal{L}_{\rm add}$ and the diamond surface roughness $\sigma_{DA}$ as free parameters in the fit.

\noindent For the laser-cut microdevice, the fit yields a diamond surface roughness value of \unit[0.9]{nm} and additional losses of \unit[610]{ppm}. The EBL device fit yields values of \unit[1.2]{nm} surface roughness and \unit[820]{ppm} additional losses. For both devices, a distribution of finesse values for a fixed diamond thickness is observed, leading to the band-like structure in Fig.~\ref{fig:finesse_fit}. This means that for a specific diamond thickness, lateral cavity positions with varying surface quality are found. To quantify the spread in surface roughness of the data presented in Fig.~\ref{fig:finesse_fit}, we use the following procedure: the finesse values for a thickness segment of \unit[10]{nm} are binned into a histogram, which is fitted with a Gaussian function. This fit determines for every thickness segment a mean value and a standard deviation. We fit the mean plus (minus) the standard deviation values of all segments with the model losses of equation~(\ref{equ:effective_losses}), to determine the upper yellow (lower bright green) curves in Fig.~\ref{fig:finesse_fit}. With that, we find that our laser-cut (EBL) device exhibits areas with surface roughness values as low as \unit[0.6]{nm} (\unit[1.1]{nm}), which are common values for these fabrication methods. For both samples finesse values similar to the bare cavity are reached in the air-like modes, whereas in the diamond-like modes scattering losses of $\mathcal{L}_{S,\rm{eff},d}=\unit[390]{ppm}$ (\unit[1330]{ppm}) limit the achievable finesse in the laser-cut (EBL) device.

\subsection{Frequency splitting of the polarization cavity modes}

In addition to the finesse, the frequency splitting of the two orthogonal polarization modes of the microcavity is studied. Strain-induced birefringence in the diamond microdevice can cause a splitting of the horizontal and vertical polarization cavity mode~\cite{howell_strain-induced_2012}. In general, diamond color centers couple differently to these modes depending on their electric dipole overlap with the electric field of the cavity. These polarization modes are used in cross-polarization resonant excitation and detection schemes, and the magnitude of polarization splitting becomes relevant for excitation laser power considerations~\cite{yurgens_cavity-assisted_2024}.

\begin{figure}[ht]
    \centering
    \includegraphics[width=0.85\linewidth]{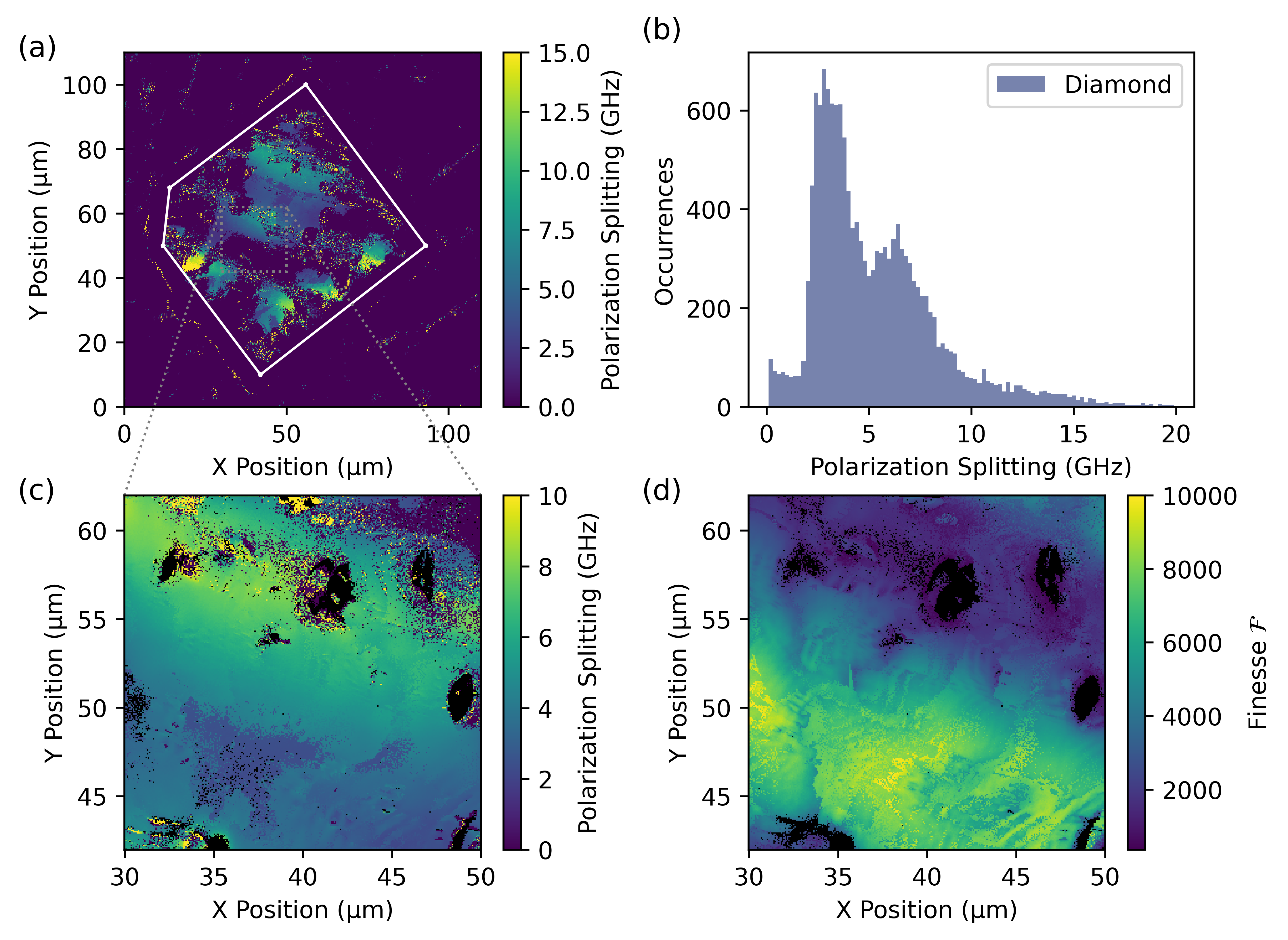}
    \caption{Polarization splitting on the laser-cut diamond microdevice (\textit{Vincent Vega}). (a) Polarization splitting is measured on the bare mirror and the diamond device. Data is taken within the same scan as shown in Fig.~\ref{fig:scm}~(a). (b) Histogram of the polarization splitting measured on the diamond device inside the white rectangle in (a). Around half of the measured lateral points do not show a polarization splitting (zero bin not included in the histogram). The bin size is \unit[0.2]{GHz}. (c) Polarization splitting, seperately measured with a \unit[70]{nm} resolution in the \unit[($20 \times 20$)]{\textmu m$^2$} area indicated in (a). (d) Corresponding finesse values of the same area as shown in~(c). The air-like modes with a higher finesse show a smaller polarization splitting compared to the diamond-like modes.}
    \label{fig:pol_scan}
\end{figure}

\noindent We investigate the polarization splitting of the fundamental modes within the same scan used for the finesse measurements. To measure the splitting, the cavity is resonantly probed using laser light with frequency-modulated sidebands, generated by a phase electro-optic modulator. The cavity length is scanned around the fundamental mode, and a typical measurement is shown in \ref{app:pol_split}. The spacing of the sidebands is set by the modulator driving frequency and is used to determine the frequency splitting of the polarization modes.\\
Figure~\ref{fig:pol_scan}~(a) shows the polarization splitting measured on and around the laser-cut microdevice. On the bare mirror, no polarization splitting is measured, while on the diamond device, polarization splittings up to \unit[15]{GHz} can be observed (histogram shown in Fig.~\ref{fig:pol_scan}~(b)). It seems that regions with higher polarization splitting in Fig.~\ref{fig:pol_scan}~(a) are closer to the device edges, which can be explained by higher strain, possibly induced by the damage of the laser-cutting. Figure~\ref{fig:pol_scan}~(c) shows a selected part of the device with the corresponding finesse scan in Fig.~\ref{fig:pol_scan}~(d). It is observed that the polarization splitting is higher in the low-finesse, diamond-like mode part (upper area in the scan). This can be explained by the higher electric field intensity inside the diamond for diamond-like modes, leading to a more pronounced effect of the birefringence~\cite{korber_scanning_2023}.\\
For the EBL microdevices presented here, we do not observe significant polarization splitting beyond the cavity linewidth. However, we have also seen polarization splittings up to \unit[10]{GHz} on other EBL-fabricated devices. To understand the origin of the different strain values, further investigations are needed.

\section{Optically coherent color centers}

\begin{figure}[ht]
    \centering
    \includegraphics[width=0.92\linewidth]{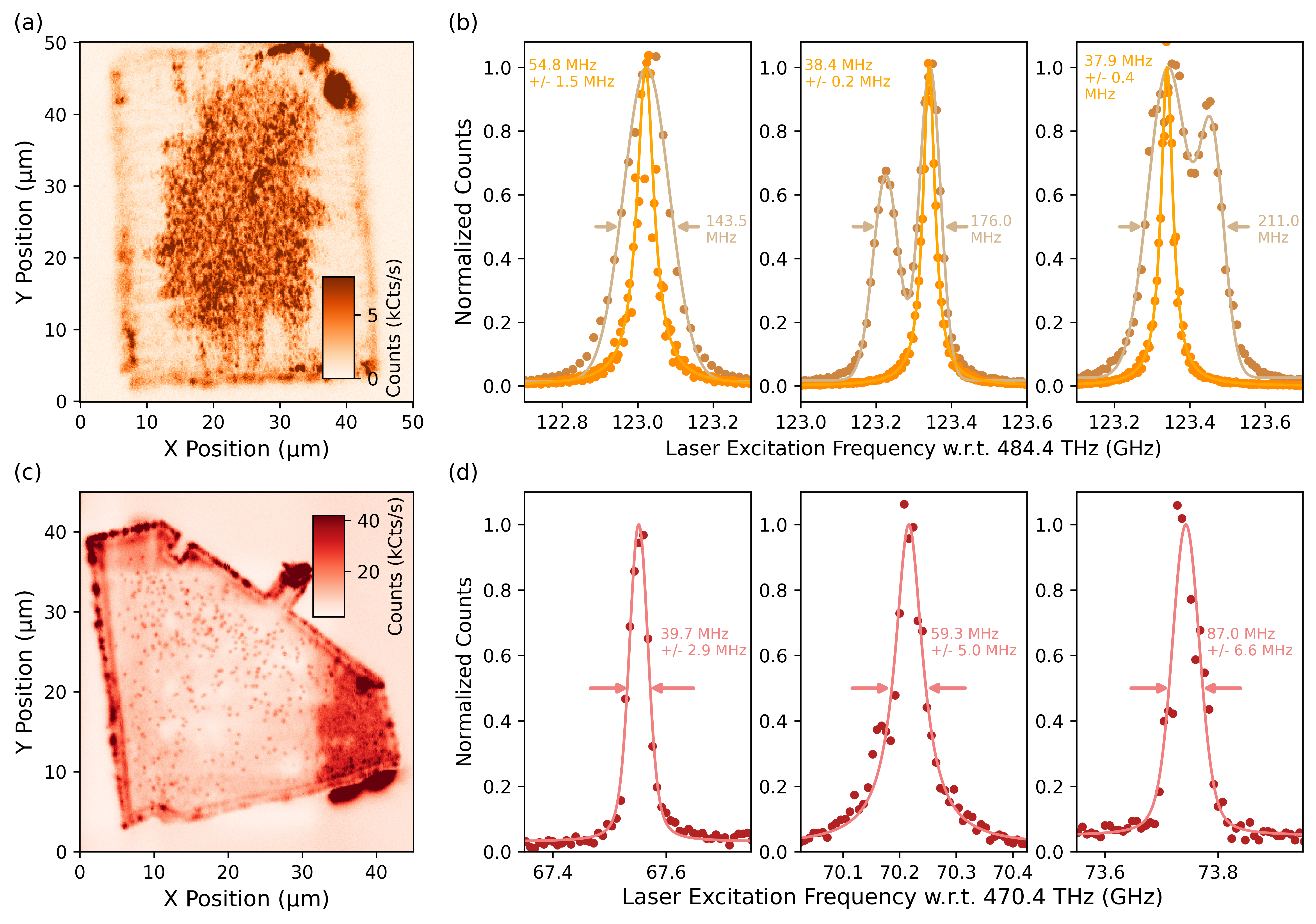}
    \caption{Optical properties of SnV centers inside a laser-cut diamond microdevice (\textit{Vincent Vega}) and NV centers inside an EBL patterned diamond device (\textit{Mr.~Orange}). The emitter linewidth measurements are performed at $\sim\unit[4]{K}$ with a confocal microscope. (a) A map of the SnV device with off-resonant, \unit[515]{nm} excitation. The fluorescence is filtered with a $\unit[620]{nm}$ bandpass filter (full-width half maximum of $\unit[10]{nm}$) to detect predominantly fluorescence of SnV centers. A high SnV center density is observed in the center of the device, while the SnV center layer is removed at the wedged edges. Note that this map was measured in a room-temperature scanning confocal microscope. (b) Typical PLE laserscans of SnV centers, fitted with a Gaussian (ocher) and Lorentzian (orange) function to obtain the spectral diffusion and dephasing linewidth, respectively. (c) Confocal map of the electron-beam patterned device, showing individual, spatially separated NV centers. (d) Typical PLE laser scans of NV centers, fitted with a Voigt function to extract the linewidth. To initialize the NV center into the bright state, a green laser pulse is applied in every repetition. The relevant measurement parameters for the scans in (b) and (d) can be found in \ref{app:ple}.}
    \label{fig:ple}
\end{figure}

For quantum network applications, the color centers in the diamond microdevices must show a (close to) transform-limited optical emitter linewidth. Noise in the environment of the color center (such as fluctuating charges) broadens the emitter linewidth. We can decompose the linewidth broadening into a slow noise contribution, spectral diffusion, and a fast noise contribution, pure dephasing~\cite{beukers_remote-entanglement_2024}. The former leads to a Gaussian broadening and can be mitigated by active tracking and feedback on the emitter transition frequency~\cite{hensen_loophole-free_2015,brevoord_heralded_2024,brevoord_large-range_2025,van_de_stolpe_check-probe_2025}. Whereas pure dephasing leads to a Lorentzian broadening with a more fundamental influence, as it directly determines the indistinguishability in a two-photon quantum interference experiment of color centers~\cite{bernien_two-photon_2012,sipahigil_indistinguishable_2014} and limits the achievable remote entanglement fidelity in emission-based entanglement protocols~\cite{bernien_heralded_2013, beukers_remote-entanglement_2024}. Moreover, the pure dephasing linewidth enters the coherent cooperativity determining fidelities of quantum information protocols~\cite{borregaard_quantum_2019}.\\
We study the optical emitter linewidth of the color centers in our diamond microdevices with photoluminescence excitation (PLE) measurements. These measurements are performed in a home-built confocal microscope setup, where the samples are mounted in a closed-cycle Helium cryostat. In the PLE measurements, a resonant laser is scanned over the zero-phonon line (ZPL) of the color center, and the fluorescence is detected in the phonon sideband.\\
Figure~\ref{fig:ple}~(a) shows a confocal scan of the laser-cut diamond microdevice hosting SnV centers. Note that this device is a different microdevice than shown before, but it originated from the same fabrication run and parent membrane (\textit{Vincent Vega}). For the PLE measurements, an off-resonant, \unit[515]{nm} laser is used before each resonant laser scan to initialize the SnV center in its negatively charged state. The application of a green laser pulse is known to change the charge environment of the color center, resulting in spectral diffusion.
We acquire \unit[100]{} of these scans on three different SnV centers (Fig.~\ref{fig:ple}~(b)) and average all scans to get a measure of the spectral diffusion linewidth. In some of these measurements, a bistability of the SnV center transition frequency is observed. These discrete jumps between two spectral positions separated by about \unit[100]{MHz} might be caused by a charge trap near the SnV center~\cite{pieplow_quantum_2024,li_atomic_2024}.\\
To quantify the pure dephasing linewidth, scans that show a complete Lorentzian fluorescence peak are manually selected (more than 50~\%), fitted, and centered before averaging. The SnV centers of Fig.~\ref{fig:ple}~(b) show good optical coherence with pure dephasing linewidths close to the transform limit around \unit[32]{MHz}. The corresponding excited state lifetime was determined in previous work~\cite{herrmann_coherent_2024}, which also demonstrated the coherent cavity coupling of a single SnV center in a device obtained from the same diamond membrane. These results demonstrate that coherent SnV centers are created in the laser-cut microdevices.\\
Figure~\ref{fig:ple}~(c) shows the confocal scan of an EBL  diamond microdevice hosting NV centers. Note that this device is fabricated in a second run with the EBL method (device originated from parent membrane named \textit{Mr.~Orange}). For NV centers, we probe the spin-conserving $\rm E_x$ or $\rm E_y$ emitter transition of the ZPL, which involves the application of intermediate off-resonant, \unit[515]{nm}, laser pulses at every resonant laser scan frequency to mitigate spin state pumping. These off-resonant laser pulses initialize the NV center predominantly in the negatively charged and spin ground state~\cite{irber_robust_2021}.
Because of the intermediate application of green laser pulses, we can experimentally not measure spectral diffusion and pure dephasing linewidth separately. We thus analyze the emitter fluorescence peaks by fitting a Voigt profile, which is a convolution of a Gaussian and a Lorentzian function. In the fit routine, the Lorentzian linewidth is lower bound to the transform-limited linewidth of \unit[13]{MHz} of the NV center~\cite{hermans_entangling_2023}. Typical scan results with their corresponding Voigt fits are shown in Fig.~\ref{fig:ple}~(d).\\
We analyze 20 different NV centers at several lateral diamond microdevices positions and find linewidths between \unit[38]{MHz} and \unit[130]{MHz} with a median linewidth of \unit[62]{MHz} (statistics shown in Appendix Fig.~\ref{fig:nv_stats}). These results are in line with former work with larger $\unit[(2 \times 2)]{mm^2}$ thinned-down diamond membranes~\cite{ruf_optically_2019}, indicating that the much tighter lateral dimensions do not induce additional optical decoherence.

\clearpage

\section{Conclusion}

We present a method for patterning high-quality micrometer-thin diamond devices utilizing laser-cutting. This method significantly simplifies the fabrication process compared to previously reported methods. Also, the process step of laser-cutting can be readily outsourced. We show that these microdevices can be successfully bonded to a dielectric cavity Bragg sample mirror using a micromanipulator. SCM scans reveal a high surface quality of the fabricated devices, reflected by high cavity finesse values. In addition, we demonstrate that the devices host coherent color centers, which are suitable for quantum optical experiments and applications.\\
For quantum spin-photon interfaces, the expected microdevice performance in microcavities can be estimated with the measured finesse of 9000 (2000) in the air-like (diamond-like) mode. With the \unit[17.3]{\textmu m} radius of curvature of the used fiber mirror, these values translate to a maximum Purcell factor of about 30 (39) for a NV center and, with adjusted mirror coatings, also for a SnV center. The Purcell enhancement would be combined with a high outcoupling efficiency through the sample mirror of 40~\% (51~\%) for the air-like (diamond-like) modes and the used mirror coatings. The performance values could be further improved by using a fiber tip mirror that shows lower additional losses~\cite{hunger_fiber_2010,uphoff_frequency_2015,maier_fabrication_2025}.\\
Combined with the integrated gold striplines on the sample mirror for microwave control of the spin qubits~\cite{bogdanovic_robust_2017,guo_microwave-based_2023,herrmann_low-temperature_2024}, the presented approach can enable the realization of an efficient spin-photon interface with diamond color centers.

\section*{Data availability statement}

The datasets of this study and the Python software for analysis and plotting are publicly available on 4TU.ResearchData under Ref.~\cite{herrmann_data_2025}.

\ack

The authors thank Alexander Stramma for proofreading the manuscript. Furthermore, the authors thank the Quentin Tarantino movie universe for providing the sample names.\\
The authors acknowledge financial support from the Dutch Research Council (NWO) through the Spinoza prize 2019 (project number SPI 63-264) and from the Dutch Ministry of Economic Affairs and Climate Policy (EZK), as part of the Quantum Delta NL programme. The authors gratefully acknowledge that this work was partially supported by the joint research program “Modular quantum computers” by Fujitsu Limited and Delft University of Technology, co-funded by the Netherlands Enterprise Agency under project number PPS2007. This research was supported by the Early Research Program of the Netherlands Organisation for Applied Scientiﬁc Research (TNO) and by the Top Sector High Tech Systems and Materials.\\

\newpage
\noindent\textbf{Competing interests:} The authors Y.~H., J.~M.~B., J.~F., C.~S., M.~R., N.~d.~J., and R.~H. filed a patent for the method of laser-cutting device fabrication used in this study.\\
\textbf{Author contributions:} Y.~H., J.~M.~B., and J.~F. contributed equally to this work. M.~R. and N.~C. developed the EBL fabrication process used in this work. N.~d.~J. contributed to developing the laser-cutting method. J.~M.~B., C.~S., and Y.~H. fabricated the diamond devices. C.~S. and L.~G.~C.~W. characterized the devices with the white light interferometer. J.~F.,~Y.~vd.~G., C.~F.~J.~W., and S.~S. built the cavity characterization setup. S.~S. measured the two-dimensional cavity scans and characterized the cavity fiber. C.~S., Y.~H., J.~M.~B., and J.~F. characterized the color centers. R.~M. fabricated the cavity fiber. J.~F. derived the used cavity loss model. Y.~H., J.~F., and S.~S. analyzed the data. Y.~H., J.~M.~B., J.~F., S.~S., and R.~H. wrote the manuscript with input from all authors. R.~H. supervised the experiments.

\appendix

\section{Outline of fabrication}

The detailed steps of both fabrication methods are summarized in table~\ref{tab:fab}.\\

\begin{table}[ht]
\fontsize{9pt}{9pt}\selectfont
\begin{tabular}[t]{p{3.5cm}|p{6cm}|p{4.5cm}}
Fabrication Step& \centering Electron-Beam Lithography (EBL) & Laser-Cutting\\
\hline
\hline
I. Cleaning                         & \multicolumn{2}{|c}{Fuming nitric acid (1)} \\
& \multicolumn{2}{|c}{\unit[10]{min} in $\rm HNO_{3}$ (65~\%) at room temperature} \\
\hline
\hline
(II. Strain Relief Etch, Implantation, Annealing) & Strain relief etch as described in subsection~\ref{sec:patterning}, followed by tri-acid clean, implantation, annealing, and a post-anneal tri-acid clean as detailed in subsection~\ref{sec:cccreation} & - \\
\hline
III. Application of                   & \centering PECVD                  & Application of PVA layer (2) \\
Protection Mask                     & \centering ${\sim\unit[320]{nm}}$ silicon nitride $\rm Si_xN_y$    &  \\
\hline
IV. Spin Coating                     & Sample mounted with PMMA495~A4 on silicon piece, resist CSAR-13 (AR-P~6200.13), \unit[3500]{rpm} (thickness of ${\sim\unit[430]{nm}}$), bake \unit[3]{min} at 150~$^{\circ}$C & - \\
& Spin coat Electra~92 (AR-PC~5090), \unit[4000]{rpm} (thickness of  ${\sim\unit[30]{nm}}$), bake \unit[2]{min} at 90~$^{\circ}$C  & \\
\hline
V. Exposure                         & \centering Electron-beam  & - \\
 & Dose of $\unit[400]{uC/cm^2}$ with \unit[3]{nm} spot size & \\
\hline
VI. Development                      & Remove Electra~92 with \unit[1]{min} deionized water, blow dry with nitrogen &  - \\
& CSAR-13 development: \unit[1]{min} in pentyl-acetate, \unit[5]{s} in ortho-xylene, \unit[1]{min} in IPA, blow dry with nitrogen & \\
\hline
VII. Patterning Hard Mask             & \centering $\rm Si_xN_y$ etch with plasma composed of $\rm CHF_3/O_2$~\cite{ruf_resonant_2021}      &  - \\
\hline
VIII. Resist Removal                   & PRS~3000 for \unit[2]{h} at 80~$^{\circ}$C, followed by overnight PRS~3000 at room temperature        & - \\
\hline
IX. Cleaning                         & \centering Double Piranha clean &  - \\
& Mixture with ratio 3:1 of $\rm H_2SO_4$~(95~\%)~:~$\rm H_2O_2$~(31~\%) at 80~$^{\circ}$C for \unit[20]{min} & \\
\hline
X. Transfer Pattern                & \centering Anisotropic ICP/RIE  &  Laser-cutting (3) with a \\
into Diamond & Sample mounted with PMMA495~A4 on fused quartz carrier wafer, \unit[30]{min} $\rm O_2$ plasma chemistry & femtosecond pulsed laser \\
\hline
XI. Cleaning                         & \centering Removal of the $\rm Si_xN_y$ hard mask: & Removal of PVA \\
& Hydrofluoric (HF) acid (40~\%) clean \unit[15]{min} at room temperature & Ultrasonic bath in de-ionized water and acetone for \unit[10]{min}, each at room temperature (4), followed by Piranha clean \\
\hline
(XII. Strain Relief Etch, Implantation, Annealing)                   & - & Strain relief etch (5) as described in subsection~\ref{sec:patterning}, followed by tri-acid clean, implantation (6), annealing (7), and post-anneal tri-acid clean as detailed in subsection~\ref{sec:cccreation}\\
\hline
\hline
XIII. Release                 & \multicolumn{2}{|c}{ICP/RIE (8)}   \\
Etch & \multicolumn{2}{|c}{Membrane protected by fused quartz mask with central opening, 45 min Ar/$\rm Cl_2$} \\
& \multicolumn{2}{|c}{Multiple rounds of $\rm O_2$ etch until individual microdevices are released.} \\
& \multicolumn{2}{|c}{Etch rates can be found in Ref.~\cite{ruf_optically_2019}}.\\
\hline
XIV. Cleaning                         & \multicolumn{2}{|c}{HF clean}  \\
\hline
XV. Bonding to Mirror                & \multicolumn{2}{|c}{Micromanipulator (9)} \\
\end{tabular}
\caption{Detailed fabrication steps and process parameters of the two fabrication methods to obtain diamond microdevices, starting with $\sim\unit[50]{\upmu m}$ membranes. All wet processing steps are followed by a dip in acetone and IPA and blow-drying with a nitrogen gun. The numbers in brackets for the laser-cutting refer to Fig.~\ref{fig:scheme}~(c) of the main text. Fabrication step II. (XII.) of the EBL (laser-cutting) method is used to create SnV centers. In the case of NV centers, the color centers are created in the bulk diamond, before slicing into membranes, as described in subsection~\ref{sec:cccreation}.}
\label{tab:fab}
\end{table}

\clearpage
\section{Bonding of microdevices to cavity mirror}

\begin{figure}[ht]
    \centering
    \includegraphics[width=0.91\linewidth]{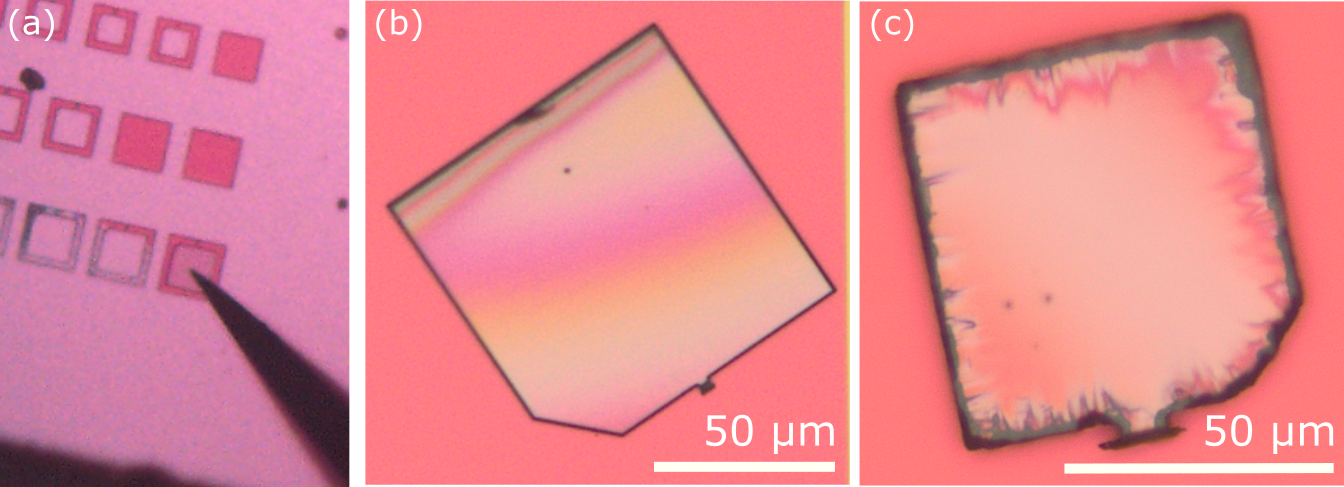}
    \caption{Bonding of diamond microdevices to the cavity mirror. (a) Released, thinned-down diamond devices above the cavity mirror. Individual devices can be broken out with the tip of a micromanipulator. An ill-bonded device fabricated by EBL (\textit{Pai Mei}) in (b) and by laser-cutting (\textit{Vincent Vega}) in (c). Both devices show interference fringes and an opaque color.} 
    \label{fig:bonding}
\end{figure}

Figure~\ref{fig:bonding}~(a) shows the breaking out of individual microdevices by a micromanipulator. Supplementary Video~1 shows the process with devices from a second laser-cut membrane. We observe that up to 30~\% of the devices bond fully to the cavity mirror. Figure~\ref{fig:bonding}~(b) and (c) show diamond devices that are not fully bonded to the mirror. When approached by the cavity fiber, such devices can be moved or picked up accidentally. 

\section{Experimental setup: room temperature microcavity}
\label{app:setup}

\begin{figure}[ht]
    \centering
    \includegraphics[width=\linewidth]{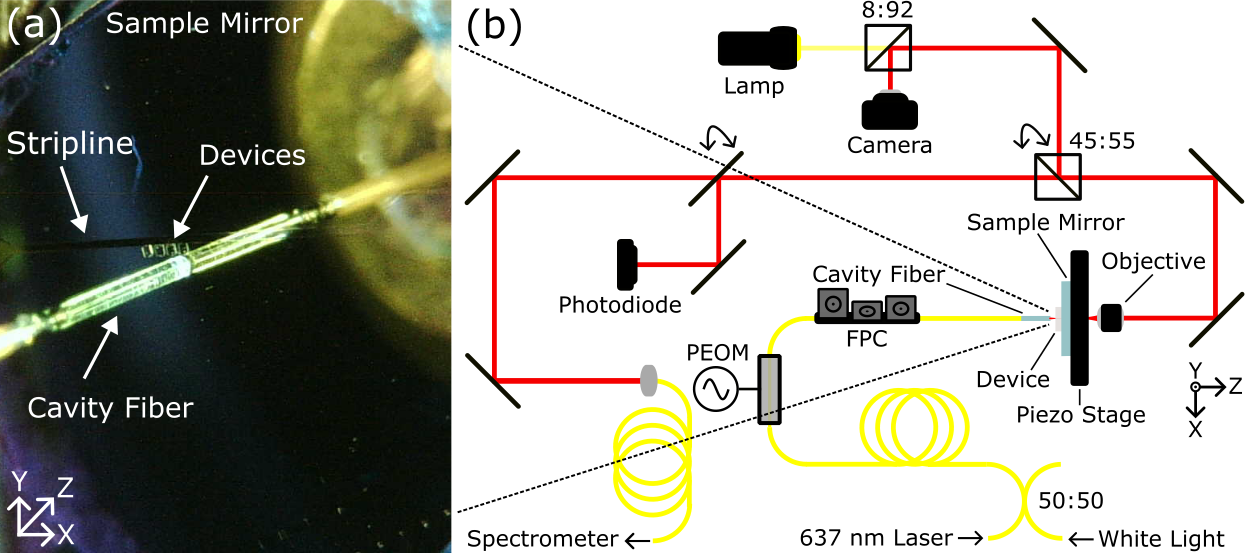}
    \caption{(a) Microscope picture of the microcavity with exemplary microdevices on the sample mirror next to a gold stripline. The cavity fiber can be seen on the left side, while the reflection in the mirror is on the right side. (b) Sketch of the home-built SCM setup. Excitation light (resonant laser or white light) is sent into the cavity via the fiber. The light can be modulated with a phase electro-optic modulator (PEOM), and the polarization is set manually with a 3-paddle fiber polarization controller (FPC). The outcoupled detection beam is collimated with an objective and sent either to a photodiode or coupled into a single-mode fiber for the spectrometer.} 
    \label{fig:setup}
\end{figure}

The cavity sample mirror with the microdevices (Fig.~\ref{fig:setup}~(a)) is mounted on a piezo nanopositioning stage (Physik~Instrumente~P-542.2SL) with a range of \unit[200]{\textmu m} in both the $X$- and $Y$-direction. The other cavity mirror is a coated, laser-ablated fiber tip~\cite{hunger_fiber_2010}, which can be moved in the axial direction by a high-precision objective scanner (Physik~Instrumente~P-721) to change the cavity length along the $Z$-direction.\\
Figure~\ref{fig:setup}~(b) shows a sketch of the optical setup and the microcavity. The cavity can be probed with resonant \unit[637]{nm} light (Newport~New~Focus Velocity~TLB-6300-LN) or with a white light supercontinuum source (NKT~Photonics~SC-450-2), spectrally filtered for \unit[600]{nm} to \unit[700]{nm}. The excitation light is combined in fiber and sent via the fiber side into the cavity. The transmitted light is collimated with an objective (100X~Mitutoyo~Plan~Apochromat~Objective) and collected on a photodiode (Thorlabs~APD130A2) or sent via fiber to a spectrometer (Princeton~Instruments~SP-2500i). In addition, the microcavity can be imaged from the sample mirror side by a lamp and a camera for lateral alignment. Supplementary Video~2 shows a two-dimensional cavity finesse scan over an exemplary device recorded with the camera.\\

\section{Experimental methods: finesse measurements and cavity fiber properties}
\label{app:finesse}

\begin{figure}[ht]
    \centering
    \includegraphics[width=\linewidth]{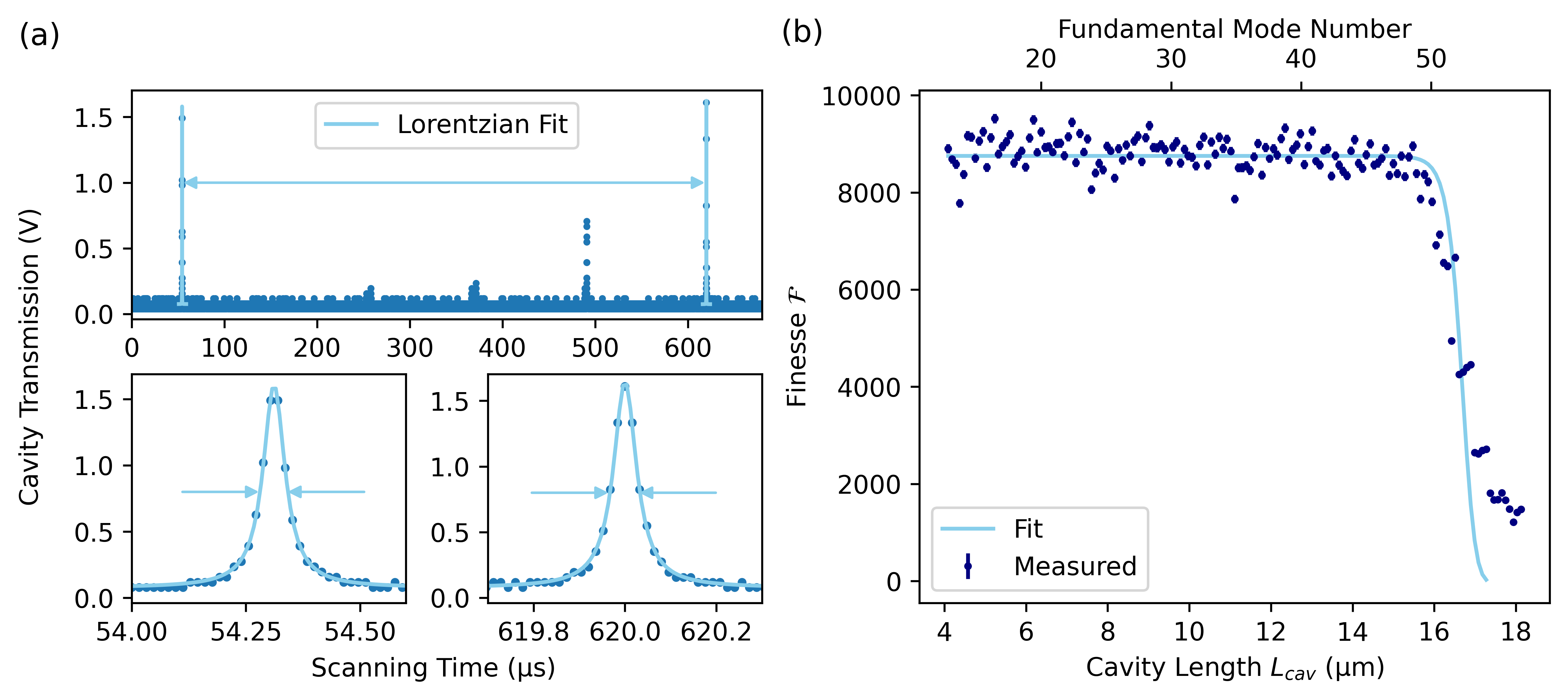}
    \caption{(a) Typical cavity finesse measurement recorded with an oscilloscope. The cavity length is scanned with a triangular voltage and two fundamental modes with a few higher-order modes are visible in the upper panel. The peaks of the fundamental modes are fitted with a Lorentzian function, yielding the cavity linewidth in time. The free spectral range in time is given by the spacing of the fundamental modes. From the shown measurement, a finesse of \unit[9500] is determined. (b) Cavity length-dependent measured finesse of the fiber used to characterize the diamond microdevices. The coating limited finesse is \unit[19000] with \unit[330]{ppm} losses. A clear plateau of finesse values of $\sim\unit[8700]{}$ is reached for a cavity length smaller than \unit[15]{\textmu m}.}
    \label{fig:fiber}
\end{figure}

To measure the finesse, the cavity is probed with a resonant \unit[637]{nm} laser, and the transmission signal is recorded on the photodiode. We record the time trace of the photodiode voltage with an oscilloscope (Picotech Picoscope 3403). The cavity length is scanned by a triangular voltage with a frequency of \unit[300]{Hz} over approximately five fundamental modes to ensure that the middle of these modes is not affected by nonlinearities occurring at the edges of the scanning range. The middle modes are fitted with Lorentzian functions, resulting in a measured cavity finesse defined as the ratio of the mode distance to the linewidth in time. An exemplary trace with two fundamental modes is shown in Fig.~\ref{fig:fiber}~(a). The resonance peak with the higher coefficient of determination ($R^2$) is used to determine the finesse. The data is filtered by requiring the mode distance to be within an acceptance range to ensure that only the fundamental modes are fitted.\\
As discussed in the main text, birefringence in diamond microdevices can cause polarization splitting, which is observed as a frequency splitting of the fundamental cavity mode. To correctly fit these modes in the finesse measurements, two Lorentzian functions are fitted (see Fig. \ref{fig:pol_measurement} (a)).\\
The same cavity fiber is used for all the presented measurements. From an interferometric measurement of the concave fiber tip, we extract a spherical radius of curvature $ROC$ of \unit[17.3]{\textmu m} with an asymmetry of 7.8~\%. To determine the range in which the fiber exhibits a stable finesse, we measure the cavity length-dependent finesse in Fig.~\ref{fig:fiber}~(b) on the bare mirror. The cavity length is determined by taking white light transmission spectra, which directly yield the cavity free spectral range and thereby the cavity length. For longer cavity lengths $L_{cav}$ clipping losses $\mathcal{L}_{clip}$ emerge which dependent on the diameter of the concave feature $D_d$ and the $ROC$ of the fiber mirror and can be calculated by~\cite{van_dam_optimal_2018}:

\begin{equation}
    \label{equ:lclip}
    \mathcal{L}_{clip} = e^{\left(-2\left(\frac{D_d}{2\omega_m}\right)^2\right)}.
\end{equation}

In this formula, $\omega_m$ denotes the cavity beam width on the fiber mirror and can be calculated by the beam waist on the sample cavity mirror $\omega_0$~\cite{van_dam_optical_2019}:

\begin{equation}
    \omega_m = \omega_o\sqrt{1+\left(\frac{L_{cav}\lambda}{\pi n \omega_0^2}\right)^2}, \quad\quad\quad \omega_0 = \sqrt{\frac{\lambda}{\pi}} \; (L_{cav}(ROC-L_{cav}))^{1/4}.
\end{equation}

\newpage
\noindent We fit the measured finesse values in Fig.~\ref{fig:fiber}~(b), with the losses determined by equation~(\ref{equ:effective_losses}) plus the clipping losses $\mathcal{L}_{clip}$. Like in section~\ref{sec:finesse_measurements} of the main text, we set $t_d=\unit[0]{\upmu m}$ and leave the additional losses and the diameter of the concave feature as free parameters. The radius of curvature is set to the measured value as stated above.\\
Note that the finesse values measured on the plateau in Fig.~\ref{fig:fiber}~(b) are a bit higher than the average bare cavity finesse values due to a local optimization on the lateral spot and reveal additional losses of \unit[390]{ppm} determined by the fit.\\

\section{Experimental methods: hybrid cavity modes}
\label{app:hybrid}

\begin{figure}[ht]
    \centering
    \includegraphics[width=0.9\linewidth]{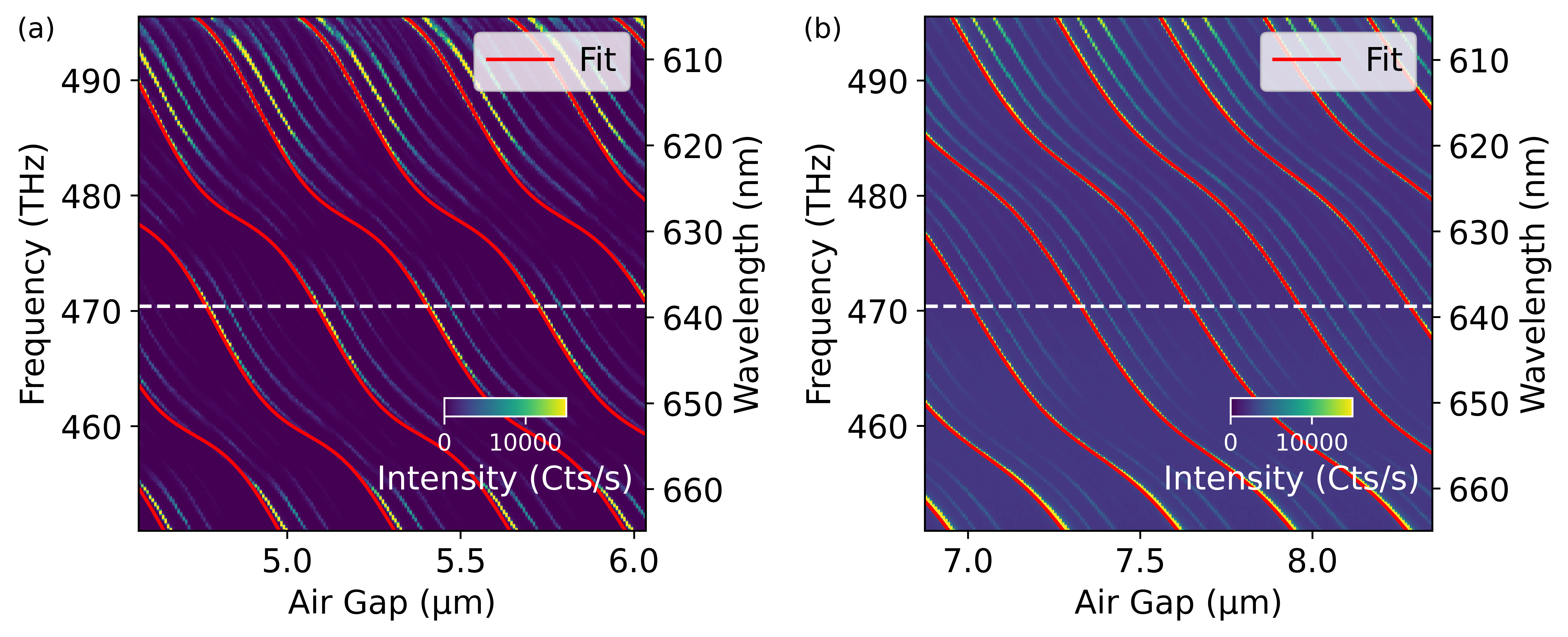}
    \caption{Cavity dispersion measured by transmission spectra of a broadband light source. (a) On the EBL microdevice (\textit{Pai Mei}), we measure a device thickness of \unit[3.31]{\textmu m}. (b) On the laser-cut microdevice (\textit{Vincent Vega}), the fit determines a thickness of \unit[2.51]{\textmu m}. Both devices are probed in the air-like mode (steeper slope) at \unit[637]{nm}, indicated by the white, dashed line. The positions on the devices of these measurements are indicated in Fig.~\ref{fig:scm}~(a) and (c) with a red cross.}
    \label{fig:wavy}
\end{figure}

To determine the cavity length and the thickness of incorporated devices, we probe the cavity with broadband white light and send the transmitted light to a spectrometer. The cavity length is changed by applying a voltage to the objective scanner, which holds the cavity fiber.\\
For the broadband white light, the cavity acts as a spectral filter, and fundamental modes appear as bright lines. We fit the fundamental modes with an analytic formula~\cite{janitz_fabry-perot_2015}, yielding the length of the air gap and the device thickness. The measurements are shown in Fig.~\ref{fig:wavy}.

\newpage
\section{Experimental methods: polarization splitting}
\label{app:pol_split}

\begin{figure}[ht]
    \centering
    \includegraphics[width=0.8\linewidth]{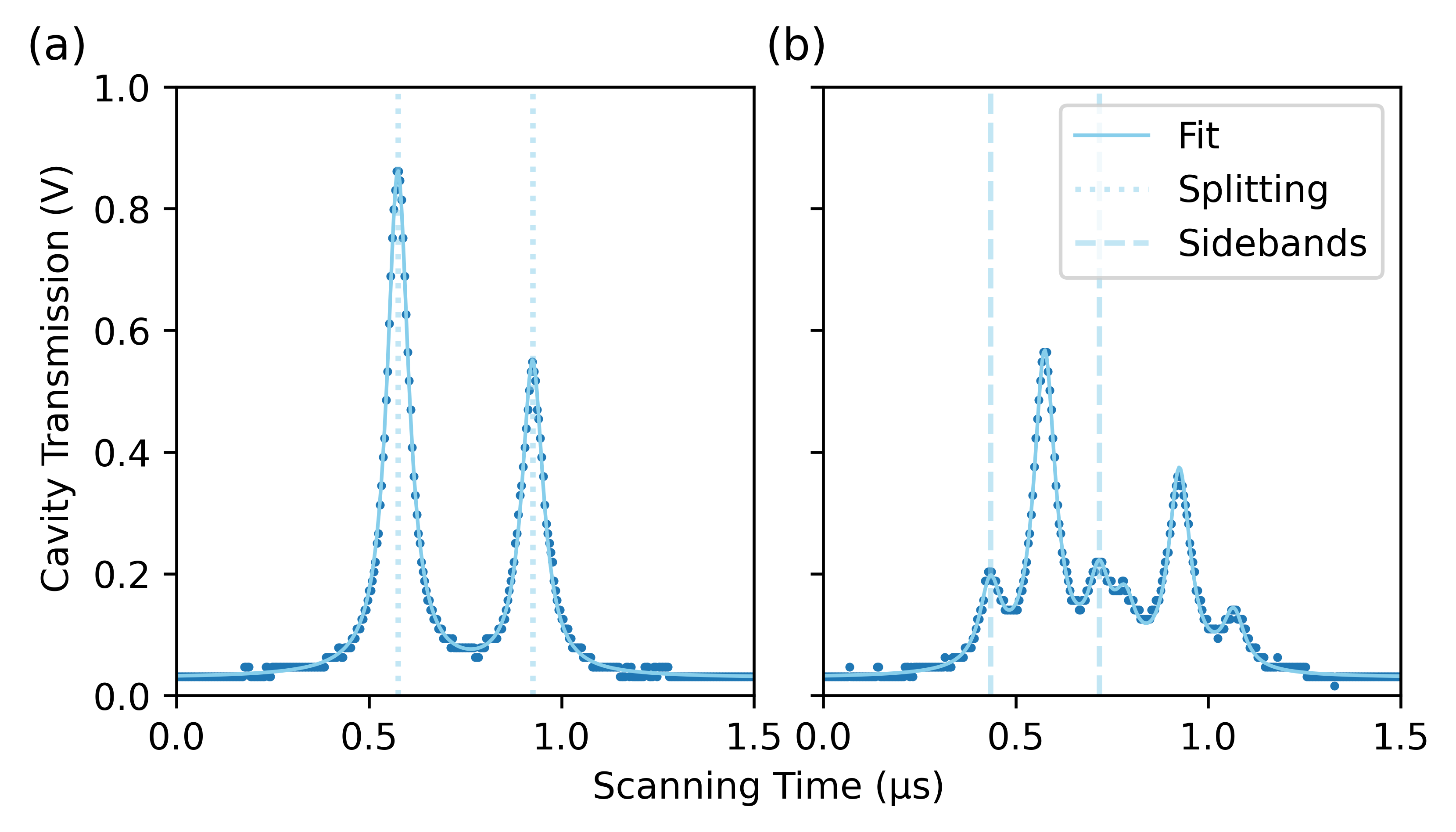}
    \caption{Exemplary measurement of the polarization splitting of the two orthogonal cavity modes on the diamond device. The cavity length is scanned with a triangular voltage over the fundamental mode, and the transmission is measured on a photodiode. (a) The length-dependent polarization splitting of the fundamental mode. (b) The same measurement with the sideband modulation. The fixed distance of the sidebands (indicated by the dashed lines) is used to determine the cavity linewidth and polarization splitting in frequency. In this measurement, a linewidth of \unit[2.9]{GHz} and polarization splitting of \unit[14.9]{GHz} (dotted lines in (a)) is extracted.} 
    \label{fig:pol_measurement}
\end{figure}

To measure the cavity polarization splitting, a second trace of the resonance peaks is measured at each lateral scan point with the resonant \unit[637]{nm} laser, modulated by sidebands. The sidebands are imprinted by a phase electro-optic modulator (Jenoptik PM635) with a microwave source (Rohde \& Schwarz SGS100A) at a frequency of \unit[6]{GHz}. The obtained modes are fitted with three (or six) Lorentzian peaks, and the sidebands are used to convert the polarization splitting from scanning time to frequency. The data are filtered by requiring $R^2 > 0.95$ to ensure the fit succeeded. An example measurement is shown in Fig.~\ref{fig:pol_measurement}.\\

\section{Experimental methods: PLE scans}
\label{app:ple}

\begin{figure}[ht]
    \centering
    \includegraphics[width=\linewidth]{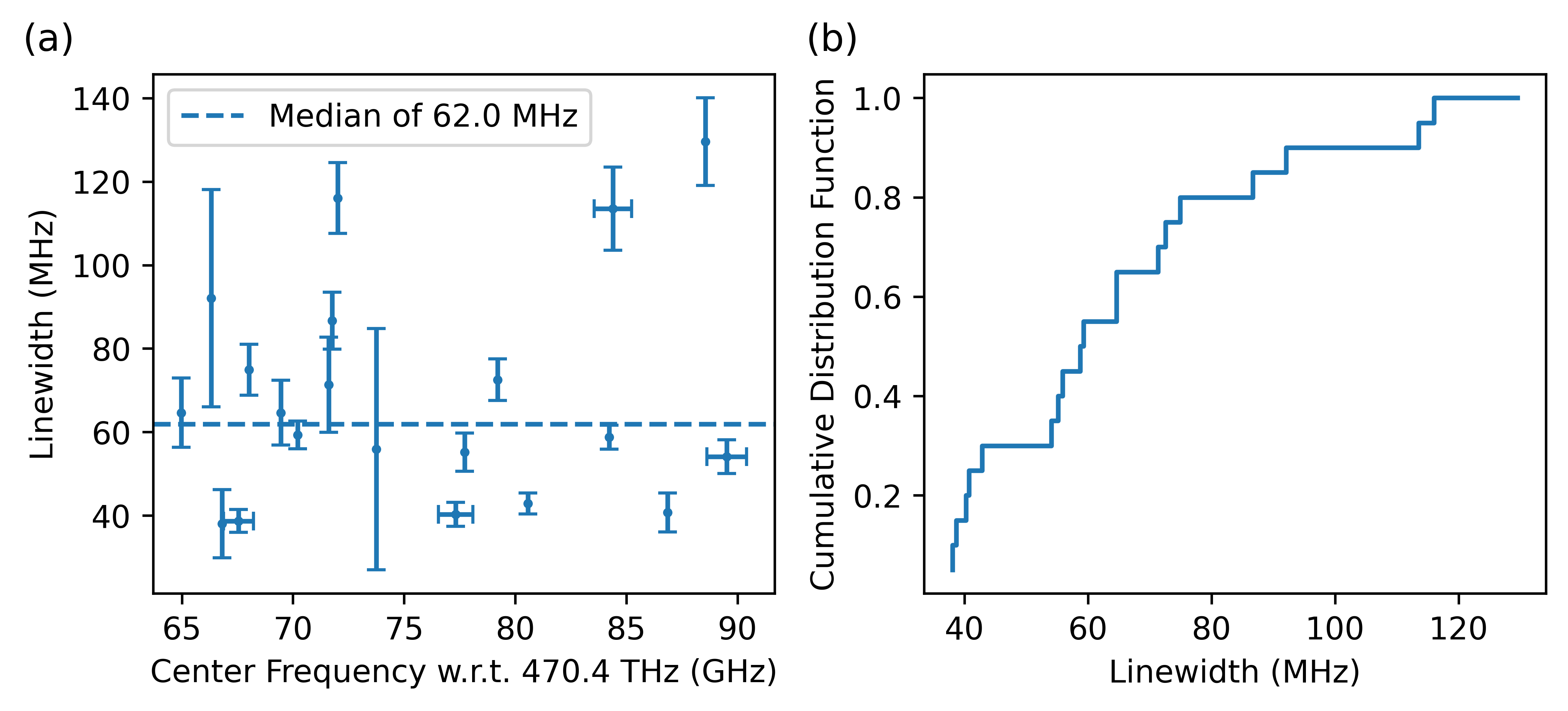}
    \caption{Statistics of the optical coherence of 20 different NV centers in an EBL patterned diamond microdevice (\textit{Mr.~Orange}), measured at $\sim\unit[4]{K}$ with a confocal microscope. (a) ZPL center frequencies and linewidth values, obtained by a Voigt fit to the individual NV center resonances (see Fig. \ref{fig:ple} for exemplary data and fit). The errors are also obtained by the fit. (b) The linewidth values are plotted as a cumulative distribution function (CDF). The measurement parameters can be found in \ref{app:ple}.}
    \label{fig:nv_stats}
\end{figure}

Measurements are performed in a home-built, cryogenic confocal microscope, whose excitation and detection path can be adapted for the measurements of SnV and NV centers. Details about the low-temperature setup and the measurement methods are described in the Supplementary Information of Ref.~\cite{brevoord_heralded_2024} and~\cite{ruf_optically_2019}.\\
The parameters for the SnV center measurements shown in Fig.~\ref{fig:ple}~(b) are: \unit[1]{\textmu W} green repump for \unit[100]{ms} and \unit[0.5]{nW} to \unit[1]{nW} resonant laser excitation power for \unit[10]{ms} integration. The resonant laser is detuned in steps between \unit[10]{MHz} to \unit[20]{MHz} with a speed of $\sim\unit[2]{GHz/s}$.\\
The parameters of the NV center measurements in Fig.~\ref{fig:ple}~(d) are: \unit[40]{\textmu W} green repump for \unit[10]{\textmu s} and \unit[40]{nW} resonant laser excitation power for \unit[20]{\textmu s} integration. At each frequency step between \unit[8]{MHz} to \unit[20]{MHz}, a sequence of repump, waiting of \unit[10]{\textmu s}, and readout is repeated for a total time of \unit[200]{ms}. The statistics of the NV center measurements can be found in Fig.~\ref{fig:nv_stats}.

\section{Measurements on the laser-cut microdevice}

\begin{figure}[ht]
    \centering
    \includegraphics[width=0.7\linewidth]{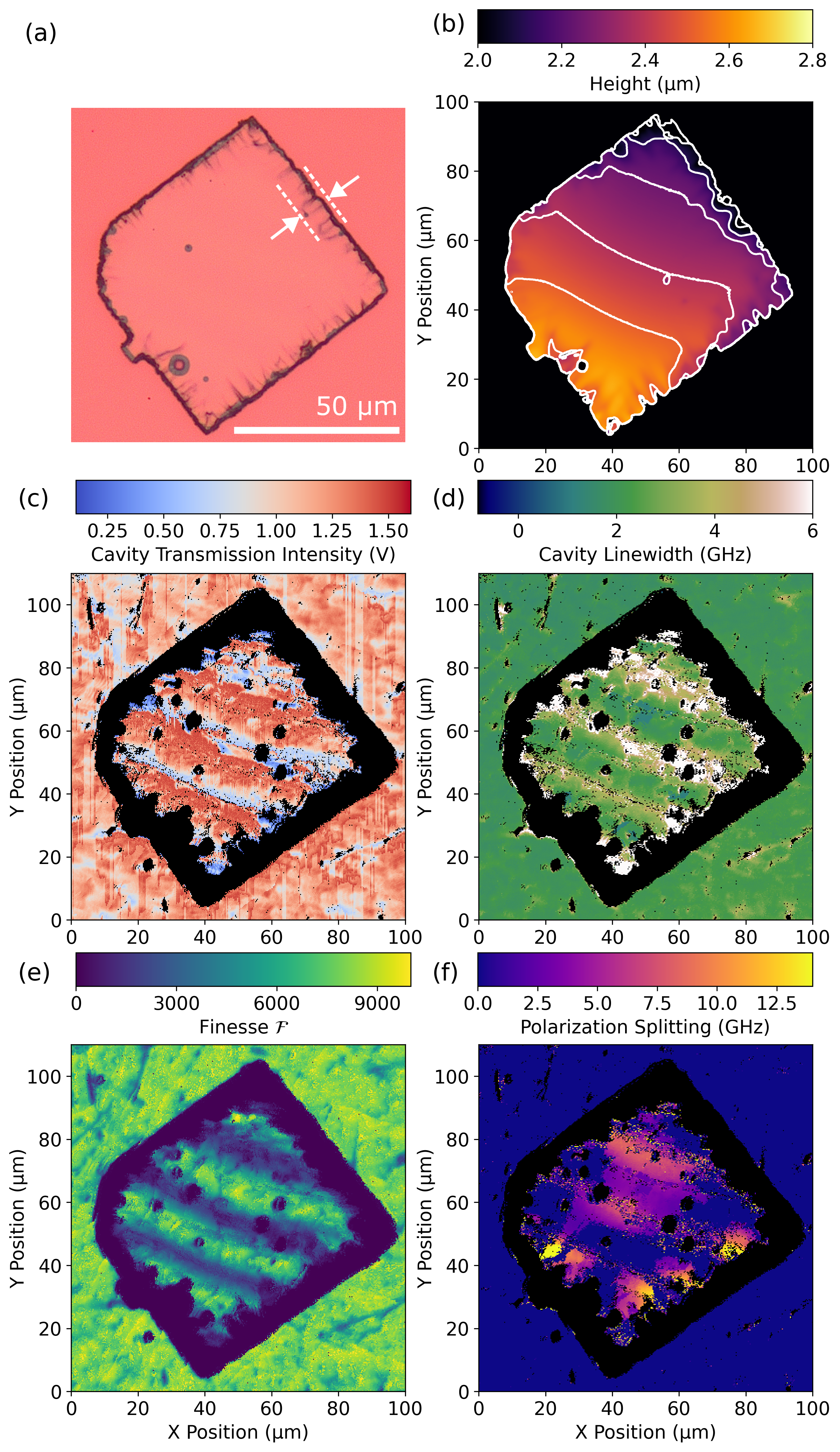}
    \caption{Summary of different measurements of the laser-cut diamond microdevice (\textit{Vincent Vega}). (a) Light microscope image. (b) Height map measured by a white light interferometer. (c)-(f) Scanning cavity microscopy. All data is measured within one scan and by probing the cavity with a resonant laser. (c) Cavity transmission intensity is measured on a photodiode. Note that the laser power is varied during the measurement. (d) Cavity linewidth. (e) Cavity finesse. (f) Cavity polarization splitting.}
    \label{fig:scans}
\end{figure}

An overview of all measurements performed on the laser-cut microdevice (\textit{Vincent Vega}) is shown in Fig.~\ref{fig:scans}.

\clearpage

\section*{References}

\bibliography{bib}

\end{document}